# Discovery of a topological excitonic insulator with tunable momentum order

**Authors:** Md Shafayat Hossain[1*†], Zi-Jia Cheng[1*], Yu-Xiao Jiang[1*], Tyler A. Cochran[1*], Song-Bo Zhang[2, 3*], Huangyu Wu[4, 5*], Xiaoxiong Liu[3], Xiquan Zheng[6], Guangming Cheng[7], Byunghoon Kim[1], Qi Zhang[1], Maksim Litskevich[1], Junyi Zhang[8], Jinjin Liu[4, 5], Jia-Xin Yin[1], Xian P. Yang[1], Jonathan D. Denlinger[9], Massimo Tallarida[10], Ji Dai[10], Elio Vescovo[11], Anil Rajapitamahuni[11], Hu Miao[12], Nan Yao[7], Anna Keselman[13, 14], Yingying Peng[6], Yugui Yao[4, 5], Zhiwei Wang[4, 5†], Luis Balicas[15], Titus Neupert[3], M. Zahid Hasan[1, 7, 16†]

**Affiliations:**

[1]Laboratory for Topological Quantum Matter and Advanced Spectroscopy, Department of Physics, Princeton University, Princeton, New Jersey, USA.

[2]Hefei National Laboratory, Hefei, China.

[3]Department of Physics, University of Zurich, Winterthurerstrasse, Zurich, Switzerland.

[4]Centre for Quantum Physics, Key Laboratory of Advanced Optoelectronic Quantum Architecture and Measurement (MOE), School of Physics, Beijing Institute of Technology, Beijing 100081, China.

[5]Beijing National Laboratory for Condensed Matter Physics and Institute of Physics, Chinese Academy of Sciences, Beijing 100190, China.

[6]International Center for Quantum Materials, School of Physics, Peking University, Beijing 100871, China.

[7]Princeton Institute for Science and Technology of Materials, Princeton University, Princeton, New Jersey 08544, USA.

[8]Institute for Quantum Matter and Department of Physics and Astronomy, Johns Hopkins University, Baltimore, Maryland 21218, USA.

[9]Advanced Light Source, Lawrence Berkeley National Laboratory.

[10]ALBA Synchrotron Light Source, Cerdanyola del Vallès, 08290 Barcelona, Spain.

[11]National Synchrotron Light Source II, Brookhaven National Laboratory, Upton, New York 11973, USA.

[12]Material Science and Technology Division, Oak Ridge National Laboratory, Oak Ridge, Tennessee 37830, USA.

[13]Department of Physics, Technion – Israel Institute of Technology, Haifa, Israel 32000

[14]The Helen Diller Quantum Center, Technion, 32000 Haifa, Israel

[15]National High Magnetic Field Laboratory, Tallahassee, Florida 32310, USA.

[16]Lawrence Berkeley National Laboratory, Berkeley, California 94720, USA.

†Corresponding authors, E-mail: mdsh@princeton.edu; zhiweiwang@bit.edu.cn; mzhasan@princeton.edu.

*These authors contributed equally to this work.

## Abstract

**Correlated topological materials often maintain a delicate balance among physical symmetries: many topological orders are symmetry protected, while most correlated phenomena arise from spontaneous symmetry breaking. It is rare to find cases where symmetry breaking induces a non-trivial topological phase. Here, we present the discovery of such a phase in $Ta_2Pd_3Te_5$, where Coulomb interactions form excitons, which condense below 100 K, opening a topological gap and creating a 'topological excitonic insulator'. Our spectroscopy reveals the full spectral bulk gap stemming from exciton condensation. This excitonic insulator state spontaneously breaks mirror symmetries but involves a very weak structural coupling, as indicated by photoemission spectroscopy, thermodynamic measurements, and a detailed structural analysis. Notably, scanning tunneling microscopy uncovers gapless boundary modes in the bulk insulating phase. Their magnetic field response, together with theoretical modeling, suggests a topological origin. These observations establish $Ta_2Pd_3Te_5$ as the first confirmed topological excitonic insulator in a three-dimensional crystal. This allows to access the associated physics through bulk-sensitive techniques. Furthermore, we uncover another surprising aspect of the topological excitonic insulator, a secondary excitonic instability near 5 K that breaks the translational symmetry. The wavevector of this state shows an unprecedented magnetic field tunability. Thus, we unveil a unique sequence of topological exciton condensations in a bulk crystal, offering new opportunities to study critical behavior and excitations.**

## Main text

In the pursuit of new collective phases of matter that exhibit intriguing quantum properties, one long-standing goal has been the realization of the excitonic insulator. This excitonic phase of matter, initially proposed in theoretical studies dating back to 1964, is an exotic state where excitons spontaneously form and undergo Bose-Einstein condensation in thermodynamic equilibrium[1-4]. This elusive phenomenon has attracted significant attention due to its predicted phenomenology, such as dissipationless energy transport akin to a superfluid[5], electronic ferroelectricity[6], and superradiant emission[7]. Formally, the excitonic insulator shares similarities with superconductors, as both involve many-body effects beyond non-interacting electron theory and the spontaneous emergence of a condensate of paired fermions. In the case of the excitonic insulator, the condensate is formed by excitons, which are bound electron-hole pairs[8, 9]. While the excitonic insulator is primarily driven by electronic interactions, it is often accompanied by a pronounced structural phase transition[3, 10], leaving many experimental signatures ambiguous. For example, among the three-dimensional materials considered as potential excitonic insulators[11-16], the most promising candidate so far, $Ta_2NiSe_5$, experiences such a prominent structural phase transition[10, 17-19]. Consequently, whether $Ta_2NiSe_5$ indeed hosts an excitonic insulator state remains a subject of debate[10]. Coversely, two-dimensional systems, such as bilayer heterostructures under high magnetic fields[20-25] and materials like $WTe_2$[26, 27], have shown indications of exciton condensation. In the case of $WTe_2$, exciton condensation is observed exclusively in the monolayer form[26, 27], while the three-dimensional $WTe_2$ becomes a metal without an energy gap[28, 29]. Here, we report the discovery of an excitonic insulator phase in a three-dimensional material $Ta_2Pd_3Te_5$, where the excitonic insulator transition is accompanied by a relatively weak (but non-zero) structural coupling that does not obscure the excitonic insulator physics. In this material, our data indicates that excitons spontaneously form and condense, developing a bulk-gapped excitonic insulator state. Remarkably, our study unveils that this correlated insulating state possesses a topological character, rendering $Ta_2Pd_3Te_5$ as the first experimentally identified topological excitonic insulator in a three-dimensional material.

In a topological excitonic insulator, the exciton condensate opens a topological gap in the single particle band spectrum. Here, excitons form due to the attractive Coulomb interaction between electrons and holes near the charge-neutral point, leading to a pairing instability of the Fermi surfaces and thus opening an energy gap. This results in a thermodynamic phase transition, corresponding to the condensation of excitons, where the system transitions from a high-temperature (semi)metal to a fully gapped low-temperature insulator characterized by a topological mean-field band structure. While it is not uncommon for electronic correlations to open a charge gap (*e.g.*, BCS superconductors or Mott insulators), it is very rare for that gap to be topologically non-trivial, as in the topological exciton insulator. In topological exciton insulators, the correlated gap is accompanied by gapless topological boundary states. Although the topological excitonic insulator state has been theoretically suggested for $Ta_2NiSe_5$[30], its presence remains inconclusive given that the experimental evidence for the excitonic insulator state in $Ta_2NiSe_5$ is itself still under debate. It is, therefore, desirable to find new materials that can host a topological excitonic insulator phase. In so doing, we demonstrate that $Ta_2Pd_3Te_5$ offers a unique competition between two symmetry-distinct topological excitonic insulator phases (that are mirror-symmetry-broken and with or without translational symmetry breaking). The accessibility of the resulting transition opens opportunities for studying critical behavior and emergent excitations that link topology and symmetry breaking. Along these lines, our spectroscopic investigations reveal the tunable nature of the translation breaking under a magnetic field.

Our investigation begins by characterizing the atomic structure of $Ta_2Pd_3Te_5$, which adopts an orthorhombic phase with the space group *Pnma* (No. 62) (Fig. 1**a**,**b**)[31]. The unit cell comprises two $Ta_2Pd_3Te_5$ monolayers that are stacked along the $a$-axis through weak van der Waals interactions. Each monolayer encompasses a Ta-Pd mixed layer sandwiched between two Te layers. To confirm the pristine atomic structure of our $Ta_2Pd_3Te_5$ sample, we conducted detailed atomic imaging on the (010) and (001) surfaces using scanning transmission electron microscopy (Extended Fig. 1). The obtained results confirm the expected lattice parameters of $Ta_2Pd_3Te_5$, with values of $a = 14.08$ Å, $b = 3.72$ Å, and $c = 18.66$ Å, which are consistent with previous investigations[31]. Elemental mapping using an energy-dispersive X-ray detector (Extended Fig. 2) demonstrates unperturbed atomic layers, verifying the structural integrity and composition of $Ta_2Pd_3Te_5$. Furthermore, selected area electron diffraction patterns acquired at temperatures of 290 K (top) and 90 K (bottom) (Extended Fig. 1) exhibit identical crystal symmetries and do not detect any structural phase transition in $Ta_2Pd_3Te_5$ within this temperature range. We also refer to Extended Fig. 3 for X-ray diffraction results elaborated on below. Atomically resolved scanning tunneling microscopy (STM) topographic measurements reveal a freshly cleaved $Ta_2Pd_3Te_5$ (100) surface comprising quasi-one-dimensional atomic chains extending along the $b$-axis (Fig. 1**c**). It corroborates the expected atomic structure. The lattice constants obtained from the STM topography are $b$ = 3.7 Å and $c$ = 18.6 Å. These values match our scanning transmission electron microscopy results ($b$ = 3.72 Å, $c$ = 18.66 Å) and the crystallographic structure model ($b$ = 3.713 Å, $c$ = 18.63 Å). Furthermore, the angle between the $b$- and $c$-axes appears to be approximately 90°, which is consistent with the crystallographic structure for $Ta_2Pd_3Te_5$ and our scanning transmission electron microscopy results.

Having identified a pristine surface, we conducted spectroscopic measurements. The spatially-averaged differential conductance (d$I$/d$V$) spectra exhibit temperature-dependent characteristics of gap formation (Fig. 1**d**). At higher temperatures ($T \gtrsim 100$ K), the system exhibits a gapless, semimetallic state, as evident from the nearly-V-shaped d$I$/d$V$ spectrum with low but non-zero local density of states at the Fermi energy. However, an insulating gap (Extended Fig. 4) emerges at $T \lesssim 100$ K. The gap gradually increases in magnitude as the sample is cooled (Fig. 1**e**). The development of this insulating energy gap can also be seen via angle-resolved photoemission spectroscopy (ARPES) by exploiting matrix-element selection effects and temperature control (Supplementary Notes 1, 2 and Supplementary Figs. 1-3).

To understand the anomalous semimetal-insulator transition, we conducted polarization-dependent ARPES measurements. This technique leverages linearly polarized X-rays to provide a symmetry fingerprint of the electronic states[32, 33]. Specifically, band hybridization between bands of distinct orbital characters is detectable through a change in the photoemission intensity for certain light polarizations. We oriented the sample such that the $\Gamma_0 - Z_0$ path lies within the scattering plane, ensuring that photoelectrons emitted from the Γ-point were free from any experimental asymmetry. We measured the energy-momentum cut along the $\Gamma_0 - Y_0$ path with *s/p* polarizations at both 330 K and 10 K, and the results are summarized in Fig. 2. Here, *s* and *p* designate light with an electric field perpendicular and parallel to the plane of incidence, respectively.

At $T$ = 330 K, the ARPES spectra measured with *p*-polarization primarily show one parabolic conduction band (CB1) and an M-shaped hole-like valence band (VB1) (Fig. 2**a**). The bottom of CB1 is located at the Γ-point and slightly below the Fermi level, leading to the (semi)metallicity of the crystal. Conversely, only a hole-like valence band (VB2) appears under the *s*-polarized probing light (Fig. 2**c**). Intriguingly, as the temperature drops down to 10 K, the top of the VB2 band descends by 60 meV, whereas changes to the VB1 band remain negligible. Additionally, a faint but discernible spectrum of the VB1 band emerges in the *s*-polarized spectra. These results rule out a rigid band shift scenario and indicate the presence of orbital hybridizations.

We note that under the experimental geometry, the $M_y$ symmetry still holds for the photoemission process. Moreover, under the free-electron final-state approximation, we can conclude that only bands with odd (even) $M_y$ parity have finite spectral weight under *s* (*p*)-polarized light, as the mirror parity of the initial-state wavefunction must be consistent with the parity of the electron dipole-interaction to have non-zero transition probabilities. As depicted in Fig. 2**e**, CB1 and VB1 possess positive $M_y$ parities, while VB2 has a negative mirror parity in the normal phase. We note that the extracted $M_y$ parities are consistent with the first-principles calculations (Fig. 2**f**). However, at $T = 10$ K, the appearance of VB1 in both *s*- and *p*-polarized spectra indicates a breaking of the $M_y$ symmetry in the insulating phase. Moreover, the pronounced shift in VB2 and CB1, when contrasted with the stability of VB1, points to hybridization at lower temperatures between the two states, suggesting a direct-coupling ($Q = 0$) interaction. This also supports the breaking of the mirror symmetry, given the different $M_y$ parities of these states in the normal phase. Additionally, the first-principles calculations reveal that VB2 and CB1 exhibit different $M_x$ parities, further suggesting the breaking of the $M_x$ symmetry at low temperatures; see Methods Section X for symmetry analysis.

While electron-lattice interactions will generically induce a structural phase transition when exciton condensation breaks the mirror symmetries, contrasting our polarization-dependent ARPES measurements with bulk scattering probes reveals that the driving mechanism is electronic instead of structural. In fact, our structural measurements (electron and X-ray diffraction) show that any symmetry-breaking signatures occurring in conjunction with the gap formation are below our detection threshold (Extended Fig. 3, Supplementary Notes 8 and 9, and Supplementary Figs. 9 and 10). This is inconsistent with the possibility of symmetry breaking driven by the lattice. Furthermore, our specific heat measurements (Supplementary Fig. 11) uncover a weak anomaly near $T = 100$ K, which is difficult to reconcile with lattice-driven structural instabilities. Lattice-driven structural phase transitions usually lead to pronounced peaks, or lambda anomalies, in the specific heat due to the first-order nature of such transitions. Collectively, these measurements indicate that any changes in the lattice degrees of freedom is a secondary effect and that the mirror symmetry breaking detected by our ARPES measurements has an electronic origin. Therefore, the semimetal to insulator thermodynamic phase transition and the resultant mirror symmetry breaking stem from the formation of an exciton condensate. This condensate lowers the energy of the system by opening a gap, as seen in both STM and ARPES measurements.

Our conclusion that an excitonic insulator state is present in $Ta_2Pd_3Te_5$ is based on detailed STM and ARPES measurements and is consistent with the findings of recent ARPES reports[34, 35]. These observations in $Ta_2Pd_3Te_5$ contrast with the scenario for $Ta_2NiSe_5$, where a pronounced structural phase transition coincides with the exciton formation. This concurrence raises questions about the origin of the insulating gap in $Ta_2NiSe_5$: is it driven by electronic interactions or the lattice instability [10, 17-19]? Recent studies indicate that the insulating gap in $Ta_2NiSe_5$, where a clear structural transition is detected, might primarily arise from structural instability instead of exciton condensation[10]. Conversely, possible structural changes in $Ta_2Pd_3Te_5$ due to exciton condensation are below our detection threshold, implying that the insulating gap in $Ta_2Pd_3Te_5$ primarily arises from electronic instability. Because electronic and structural degrees of freedom are thermodynamically coupled, the breaking of electronic symmetry necessarily implies a structural anomaly (*e.g.*, change in volume), although not necessarily a structural transition. The detection of mirror symmetry breaking in $Ta_2Pd_3Te_5$ via photoemission spectroscopy coupled to non-discernible structural changes through X-ray diffraction, electron diffraction, and thermodynamic measurements, suggest that the coupling between electronic and structural order parameters is very weak in this material. A conclusive understanding of how the excitonic insulator transition impacts phonons in $Ta_2Pd_3Te_5$ will require detailed structural investigations.

Having unveiled the spontaneous condensation of excitons into a correlated insulating state, we investigate the topological properties of this correlated state. A hallmark of topology is the presence of gapless boundary states protected by time-reversal symmetry within the insulating bulk energy gap. To explore these boundary states, we conducted real-space investigations using STM, which allows for direct, atomic-scale visualization with high spatial and energy resolution[36]. This powerful technique has proven effective in identifying topological boundary states in various quantum materials[36-48]. Upon scanning a freshly cleaved $Ta_2Pd_3Te_5$ crystal, we observed a well-defined monolayer atomic step edge along the $b$-axis, as identified by the topographic image and corresponding height profile in Fig. 3**a**. Remarkably, our spatially resolved spectroscopic imaging at $T = 5$ K revealed a pronounced edge state within the correlated insulating gap, as depicted by the differential conductance ($\mathrm{d}I/\mathrm{d}V$) maps at $V = 0$ mV and 30 mV in Fig. 3**a** (see Extended Fig. 5 for additional results). To characterize the spatial profile of the edge state, we extract a line profile from the $\mathrm{d}I/\mathrm{d}V$ map, as illustrated in Fig. 3**b**. The line profile demonstrates an exponential decay of the edge state along the crystal side, with a characteristic decay length of $r_0 \simeq 1.6$ nm, indicating an exponential localization of the edge state. The edge state exhibits a steeper decay on the vacuum side. Furthermore, energy-resolved spectroscopic measurements conducted on the step edge (Fig. 3**c**) reveal a substantial $\mathrm{d}I/\mathrm{d}V$ signal (represented by orange curves) around the Fermi energy, while the $\mathrm{d}I/\mathrm{d}V$ spectrum away from the step edge (violet curves) exhibits an insulating gap. Note that the Fermi energy spectral dip of the edge state can possibly be attributed to Tomonaga-Luttinger liquid behavior, a many-body ground state associated with one-dimensional bands, as reported for edge states in bismuthene[49] or $Bi_4Br_4$[48], for instance.

Although edge modes have been observed in $Ta_2Pd_3Te_5$[31], no clear evidence for their topological nature has been reported. To test the topological nature of the observed boundary state, we investigated its response to an external magnetic field. The magnetic field breaks the time-reversal symmetry, which protects the gapless topological edge states. When a magnetic field perpendicular to the *bc* plane is applied, we observe a significant suppression of the $\mathrm{d}I/\mathrm{d}V$ measured at the step edge. The field-dependent tunneling spectra, shown for $B = 2$ T, 3 T, and 6 T in Fig. 3**c**, demonstrate a gradual formation of an insulating gap at the edge state. (The field-induced suppression of the edge state is also visualized in the d$I$/d$V$ map under $B = 6$ T, depicted in Fig. 3**d**.) The emergence of this gap at the step edge bears resemblance to a Zeeman gap, which typically stems from the field-induced coupling of helical edge states in a time-reversal symmetric material[50-52]. Notably, a clear and reasonably linear increase of the energy gap in both the terrace (bulk) and the step edge is observed when plotting the energy gap magnitude derived from the field-dependent tunneling spectra as a function of the magnetic field (Fig. 3**e**), thus corroborating the anticipated behavior of the Zeeman effect. Attributing the gap only to the Zeeman effect provides an estimate for the Landé $g$-factors of the bulk and edge state, yielding values of 9 and 70, respectively. It is worth noting that these g-factor estimates also include the effect of orbital magnetization. Furthermore, since the bulk energy gap is due to interactions, its magnitude may be affected by the magnetic field, leading to a non-linear dependence of the gap on $B$ that cannot be characterized by an effective $g$-factor alone. Overall, the observed edge state, initially gapless and located within the bulk insulating energy gap, undergoes a transition to a gapped state when subjected to a time-reversal symmetry breaking perturbation, providing compelling evidence for its protection by the time-reversal symmetry, a characteristic hallmark of topological boundary states[53, 54]; see Supplementary Note 15 and Supplementary Fig. 16 for additional results.

Additionally, we conducted temperature-dependent tunneling measurements to examine the behavior of the edge state as the bulk insulating gap closes above the excitonic insulator transition temperature. The results of these experiments, depicted in Fig. 3**f**, demonstrate that the insulating bulk gap and the prominent edge state observed at $T = 5$ K disappears at $T = 120$ K. At $T = 120$ K, the bulk reverts to a gapless, semimetallic state, while the edge state becomes suppressed in d$I$/d$V$ magnitude with respect to the bulk. This temperature-dependent transition of the quantum state of the bulk and edge is also evident in the spatially resolved spectroscopic maps shown in Fig. 3**g**. At $T = 5$ K, the d$I$/d$V$ maps taken at the Fermi energy display a pronounced state at the edge, and no discernible state at the bulk, whereas at $T = 120$ K, the bulk states become more prominent with respect to the edge; see Extended Fig. 5 for additional results.

The presence of a topological boundary mode within the correlated excitonic insulator energy gap indicates that the ground state of the material is a topological excitonic insulator. This is, to our knowledge, the first observation of this elusive quantum phase in a three-dimensional material. Next, we further cooled the sample to examine the excitonic state at lower temperatures. Figure 4**a** displays a large-area topographic image and corresponding d$I$/d$V$ maps of the occupied (-200 mV) and unoccupied (100 mV) sides of the insulating gap obtained at $T = 5$ K. As alluded to in earlier discussions, in this zero momentum exciton condensate, no superlattice modulations in the charge distribution are observed, and only the Bragg peaks are visible in the respective Fourier transform images (Fig. 4**a,b**). Strikingly, however, at an even lower temperature of $T = 4.2$ K, a secondary spontaneous electronic transition that *does* break translational symmetry with an incommensurate wavevector is observed, which is clearly visualized through the topographic image in Fig. 4**c**, with the corresponding Fourier transform image displaying distinct peaks denoted as $Q_{\mathrm{exc}}$. By spatially mapping the d$I$/d$V$ on both the occupied (-200 mV) and unoccupied (100 mV) sides of the energy gap, we spectroscopically visualize the resulting superlattice modulations, demonstrating spectroscopic contrast (Fig. 4**c**). The local density of states maxima (minima) at -200 mV correspond to local density of states minima (maxima) at 100 mV, indicating a reversal of the spectroscopic contrast. This is consistent with a translational symmetry breaking order. Moreover, peaks at wavevectors $Q_{\mathrm{exc}}$ obtained from the Fourier transform of the d$I$/d$V$ maps coincide with those from the Fourier transform of the topography (Fig. 4**c**). Additionally, we investigated the energy dependence of the wavevectors $Q_{\mathrm{exc}}$, and found that they appear to be energy-independent, as the $Q_{\mathrm{exc}}$ peak locations remain non-dispersive outside of the insulating gap (Fig. 4**d**). Within the insulating gap, by definition, no discernible spectroscopic signal is present. Notice that this secondary spontaneous electronic transition is accompanied by a significant increase in the spectral gap, as indicated in Fig. 1**d** by the traces collected at $T = 5$ K and 4.2 K.

To understand the nature of the translational breaking order formed within the primary excitonic insulator state, we have performed magnetic field dependence studies (Fig. 4**e**,**f** and Extended Fig. 6). In Extended Fig. 6, a series of topographic images acquired at various magnetic fields (0 T, 2 T, 4 T, and 6 T) at $T = 4.2$ K clearly reveal the presence of translational symmetry breaking at all fields. Interestingly, the corresponding wavevectors ($Q_{exc}$) continuously change with increasing magnetic field, as visualized by the Fourier transform of the topography. This field-induced evolution of $Q_{exc}$ is further evidenced by the spectroscopic imaging presented in Fig. 4**e**, revealing spectroscopic contrast in real space in both the occupied and unoccupied sides of the energy gap, with a contrast reversal between the two sides, akin to the zero-field data in Fig. 4**c**. Importantly, the $Q_{exc}$ peaks obtained from the Fourier transforms of the d$I$/d$V$ maps align with those obtained from the Fourier transform of the topography (Fig. 4**e**), both revealing markedly different $Q_{exc}$ peak positions compared to the zero-field data. The field-induced evolution of $Q_{exc}$, as summarized in Fig. 4**f** (also refer to Extended Fig. 6**f**), provides crucial insights into the nature of the ordering. At high temperatures (above $T \simeq 100$ K), $Ta_2Pd_3Te_5$ is a semimetal with a negative indirect band gap where particle- and hole-band extrema are separated by a finite, non-zero wavevector in the Brillouin zone. The primary exciton condensate is expected to inherit this characteristic wavevector in its effective (mean-field) band structure. The secondary order should then be interpreted as another exciton condensate with this non-zero wavevector, which manifests as the periodicity of the excitonic superlattice modulation in real space. The magnetic field induces a continuous shift in the particle- and hole-bands due to the Zeeman effect, which in turn causes the wavevector of the secondary excitonic order to change continuously with increasing magnetic field. This behavior is precisely observed in our experiments. Collectively, the experimental observations of an insulator-insulator transition (that is not driven by a structural instability), leading to the development of a translation symmetry breaking order (Supplementary Figs. 11 and 12), the lack of a Fermi surface in the parent state (Fig. 4), and the continuous magnetic field tunability of the $q$-vector (Extended Fig. 6), strongly support the excitonic character of the secondary order[55]. Note that the edge states are still present in the secondary order (Supplementary Fig. 17).

In semiconductors, the exciton energy typically shifts nearly linearly as a function of the magnetic field (Zeeman effect), which leads to an increase in emission peak intensity[56, 57]. However, the observed shift in wave vector (Extended Fig. 7) of the excitonic phase in $Ta_2Pd_3Te_5$ was measured deep within the finite-$q$ excitonic insulator state, which resembles an incommensurate charge density wave phase. Conventional charge density wave systems often display small Fermi surface pockets resulting from the opening of the associated gap and the resulting Brillouin zone folding. Small Fermi surface sheets are highly susceptible to Landau quantization and the Zeeman effect, being deformed by the latter to accommodate an integer number of magnetic flux quanta. As the magnetic field increases, the nesting vector is understood to shift in $k$-space to minimize the system's free energy to maintain the charge density wave stability. In $Ta_2Pd_3Te_5$, the finite momentum excitons below $T = 5$ K emerge from a fully gapped state without a Fermi surface (see d$I$/d$V$ spectra in Fig. 1**e**). Thus, the mechanism leading to the shift in the modulation vector of the $Ta_2Pd_3Te_5$'s finite-$q$ excitons likely differs from that of conventional charge density waves. Nevertheless, this observation would mark the first direct experimental demonstration of $q$-vector tuning of an excitonic insulating state by a magnetic field.

Furthermore, we have substantiated the bulk nature of the excitonic order via systematic transport experiments. Our transport experiments (Extended Fig. 7) support the overall STM observations as well as the related conclusions.

We observe clear electric threshold fields $E_{th}$ for non-linear carrier conduction, due to the sliding or phason mode of the translational symmetry breaking excitonic order[58], only below $T \cong 5$ K. Satellite peaks, associated with the finite momentum exciton condensate, become observable in the Fourier transform of the STM images precisely in this range of temperatures. The absence of both, translational symmetry breaking in STM images, and non-linear electrical conduction, confirms the absence of a translational symmetry breaking order and the zero-momentum nature of the electron-hole condensate observed above $T \cong 5$ K. Furthermore, $E_{th}$ continuously changes with the magnetic field, aligning with our STM results (see Methods section XIII for details). The two excitonic transitions appear also in the thermodynamic measurements (Supplementary Notes 10 and Supplementary Fig. 11), clearly indicating the bulk nature of the excitonic transitions.

To theoretically understand both exciton condensation orders in $Ta_2Pd_3Te_5$, we formulate a two-band minimum model on a square lattice (see Methods Section XIV and Supplementary Notes 12-14). With an appropriate choice of parameters, the model effectively mimics the low-energy band structure of monolayer $Ta_2Pd_3Te_5$ obtained through first-principles calculations (Extended Fig. 8). In its high-temperature phase, the system has no global gap, with the conduction band minimum situated near the $M$ point $\left(\boldsymbol{k} = (0, \pi)\right)$ and slightly below (by approximately 6 meV) the valence band maxima located along the $k_x = 0$ line (Fig. 5**a**). The model has an inverted band structure which is consistent with the angle resolved photoemission spectroscopy measurements. Due to the presence of a gap between the two bands at all momenta, the inverted electronic band structure of $Ta_2Pd_3Te_5$ can be characterized by a $Z_2$ Kane-Mele topological invariant in the 2D limit which is nontrivial. The $Z_2$ invariant, under time-reversal and spin conservation symmetry (as present in our model), is determined by integrating the spin Berry curvature across the first Brillouin zone. Consequently, under open boundary conditions, this system has states at the edge that are buried by the bulk continuum (Fig. 5**c**).

To account for the both excitonic instabilities, we incorporated long-range Coulomb interaction and applied Hartree-Fock mean-field approximation to the model. We initially identify that the primary excitonic order, which involves the pairing of electrons and holes with identical momentum, emerges at temperatures $T \leq 110$ K via a second-order phase transition as the temperature decreases. This leads to a substantial global bulk energy gap (Fig. 5**d**, **e, f**), which grows monotonically as the temperature is lowered (Fig. 5**g**). As the excitonic order strengthens, the direct gap between the conduction and valence bands is maintained (Fig. 5**h**). Consequently, the topological properties of the model are not changed by the primary excitonic order, which is evident in the persistence of the gapless edge states across the bulk insulating gap under open boundary conditions (Fig. 5**f**), and is consistent with the experimental findings. Under the excitonic instability, the conduction band minimum and the valence band maxima remain well separated in momentum space. This separation is what triggers the formation of a secondary excitonic order, where the ordering wave wavevector $\boldsymbol{Q}$ is equal to the momentum separation of the conduction band minimum and valence band maxima.

To model the secondary excitonic order, we used the same bare electronic structure and performed a self-consistent calculation to solve for the secondary excitonic order with finite momentum $\boldsymbol{Q}$. This finite-momentum excitonic order primarily involves large momentum transfers mediated by the Coulomb interaction. As a result, for the same interaction strength as that for the primary order, it tends to be significantly weaker and exhibits a lower critical temperature ($T_{c2} \approx 25$ K), as shown in Fig. 5**i**. Moreover, as the temperature decreases, the primary order first develops and opens the bulk gap, which suppresses the formation of the secondary order and further reduces $T_{c2}$. Therefore, while breaking translational symmetry, this secondary order does not change the system's topology (which is protected by time-reversal symmetry) and only minimally increases the bulk gap, aligning with the experimental observations (Fig. 5**j**). We further accounted for the Zeeman coupling produced by externally applied magnetic fields in our calculations. We find that this Zeeman term can smoothly change the momentum separation between the conduction band minimum and valence band maxima (Fig. 5**k**), which is consistent with the change in the ordering wave wavevector under the application of a magnetic field.

The demonstration of a topological exciton insulator opens the door to investigating the intriguing interplay between non-trivial band topology and quantum many-body effects in a new context. While the relationship between topology and electronic correlations has been intensely studied theoretically (*e.g.*, Kane-Mele-Hubbard insulator), there are very few experimentally confirmed systems where electronic topology and correlations interplay, especially for itinerant systems displaying a full topological gap. They include the anomalous quantum Hall phases of moiré systems and the topological Kondo insulators (*e.g.*, $SmB_6$)[59]. The topological excitonic insulator state in $Ta_2Pd_3Te_5$ now joins this rare group of correlated topological systems. The intertwined excitonic orders (Fig. 5**j**) and topological bands in $Ta_2Pd_3Te_5$ are unprecedented in other candidates for exciton insulators and invite integration into innovative optoelectronic/optical devices.

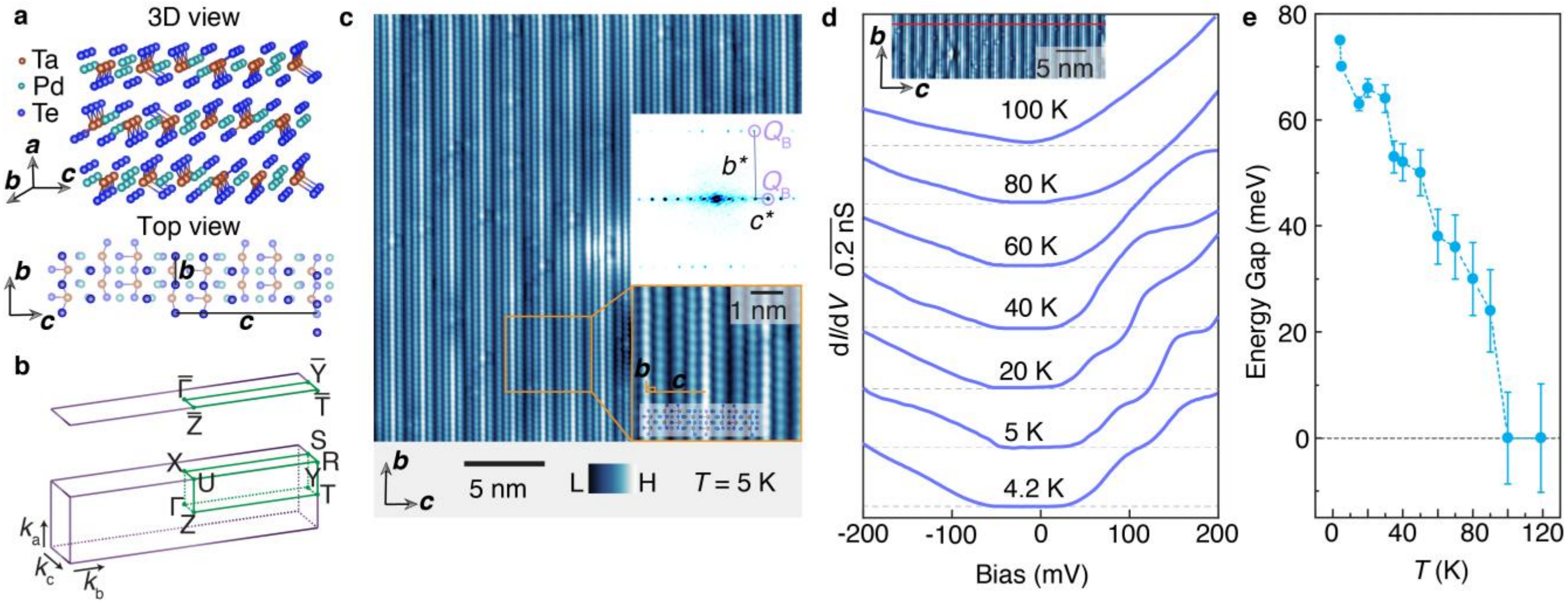

**Fig. 1: Real space characterization of $Ta_2Pd_3Te_5$ showing the development of an insulating bulk gap around $T = 100$ K.** **a**, Crystal structure (top panel) and top view of the (100) surface (bottom panel), with the highlighted Te atoms on the cleaved topmost plane of the (100) surface. The lattice constants $b$ = 3.713 Å and $c$ = 18.63 Å along the $bc$-plane are also indicated. **b**, Brillouin zone for $Ta_2Pd_3Te_5$ (bottom panel) and the surface Brillouin zone with surface normal perpendicular to the cleave plane (top panel). High symmetry points and lines are marked in green. **c**, Atomically resolved STM topographic image of the (100) plane ($V_{set}$ = 300 mV, $I_{set}$ = 0.5 nA). Bottom inset: Magnified view of the topographic image, showing lattice constants $b$ = 3.7 Å and $c$ = 18.6 Å, and an approximate 90-degree angle between them, matching the crystallographic structure. The top view of the crystal structure (shown in panel **a**) is superimposed onto the image to indicate the atomic chains. Top inset: Fourier transform of the topographic image, indicating the Bragg peaks along the $b$- and $c$- axes. **d**, Spatially averaged energy-resolved tunneling spectra obtained at various temperatures. Spectra are averaged along the red line marked in the topographic image (inset). Spectra for different temperatures are vertically offset for clarity. No gap is observed in the d$I$/d$V$ spectrum at $T = 100$ K, while at $T = 4.2$ K, a sizeable insulating gap of approximately

75 meV emerges. **e**, Temperature dependence of the insulating gap. The gap appears around 100 K and increases with decreasing temperature. Vertical bars indicate the thermal broadening energy for each data point.

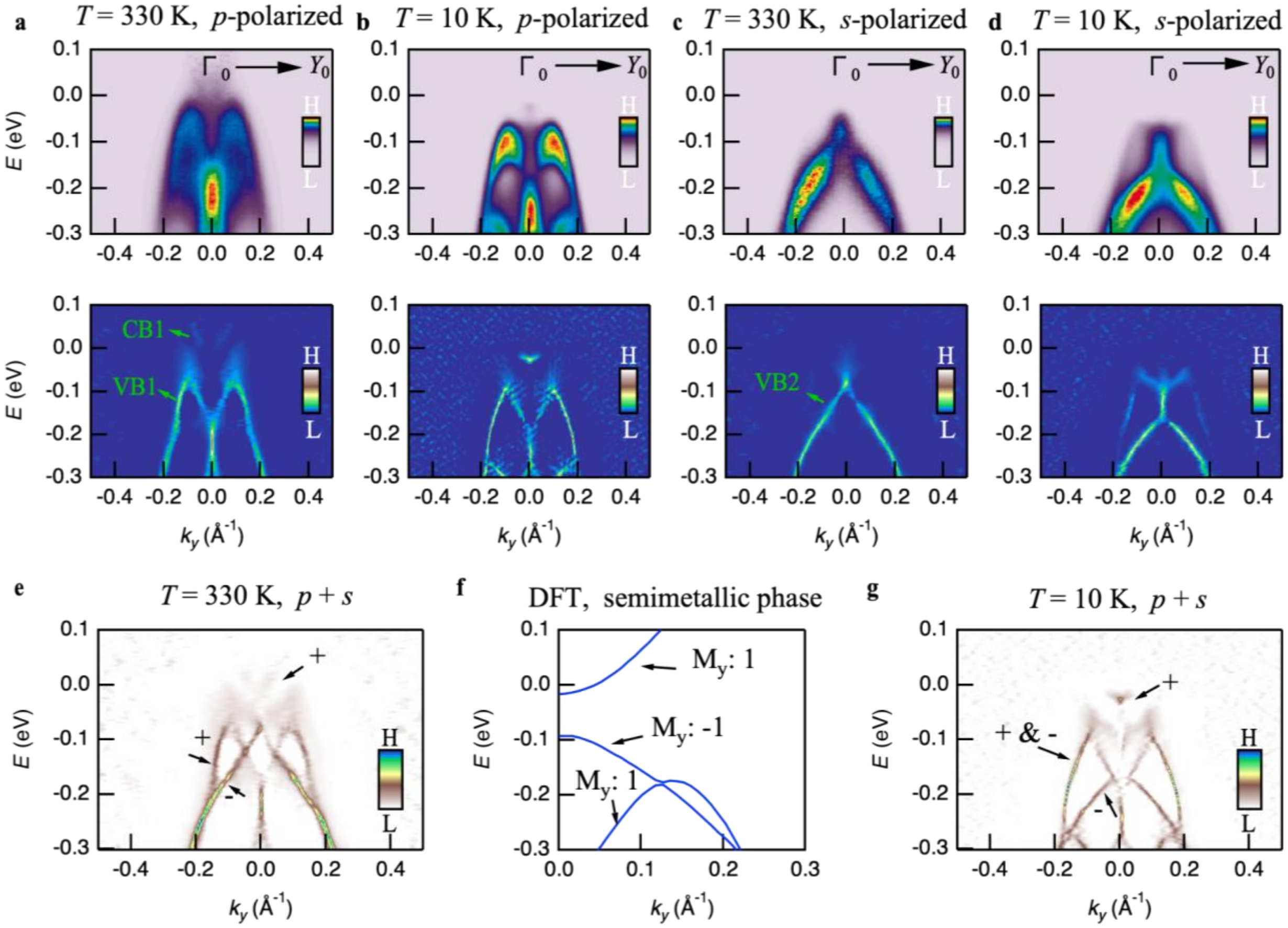

**Fig. 2: Signatures of interband hybridization and mirror symmetry breaking in the low-temperature electronic phase of $Ta_2Pd_3Te_5$. a-d,** Experimental energy-momentum dispersions along the direction, captured using 54 eV photons with *s* and *p* polarizations, at $T = 10$ and 330 K. Raw data are presented in the upper row, and the corresponding curvature plots are shown in the lower row. The conduction band (CB1) and two valence bands (VB1 and VB2) are highlighted by the green arrows. **e**, Full *E*-*k* dispersion for the normal phase, obtained by combining *s*- and *p*-polarized data. **f**, Calculated band structure for monolayer $Ta_2Pd_3Te_5$ in its semimetallic phase. The calculated mirror symmetry $M_y$ parity (without considering spin-orbit coupling) for each band is also indicated, which matches the experimental results shown in panel **e**. **g**, Same as panel **e**, but for the low-temperature excitonic insulator phase. The eigenvalues of mirror symmetry $M_y$ are extracted for each band based on their polarization dependencies and are summarized in panels **e** and **g**. Note that VB1 shows up in both *s*- and *p*-polarization channels at low temperatures, suggesting the breakdown of the $M_y$ symmetry. Meanwhile, VB2 shifts significantly down to a deeper binding energy when compared to VB1, ruling out the possibility of a rigid band shift and underscoring the presence of interaction-driven interband hybridization. Plus and minus signs indicate the mirror parities for each band. At the lowest temperatures, a distinct spectral feature emerges near the $\Gamma_0$ point in the ARPES curvature plot (panel **g**). This state possesses a positive $M_y$ character and out-of-plane dispersion and follows the spectrum intensity distribution of the CB1 at higher temperatures. (See discussion in Supplementary Note 1).

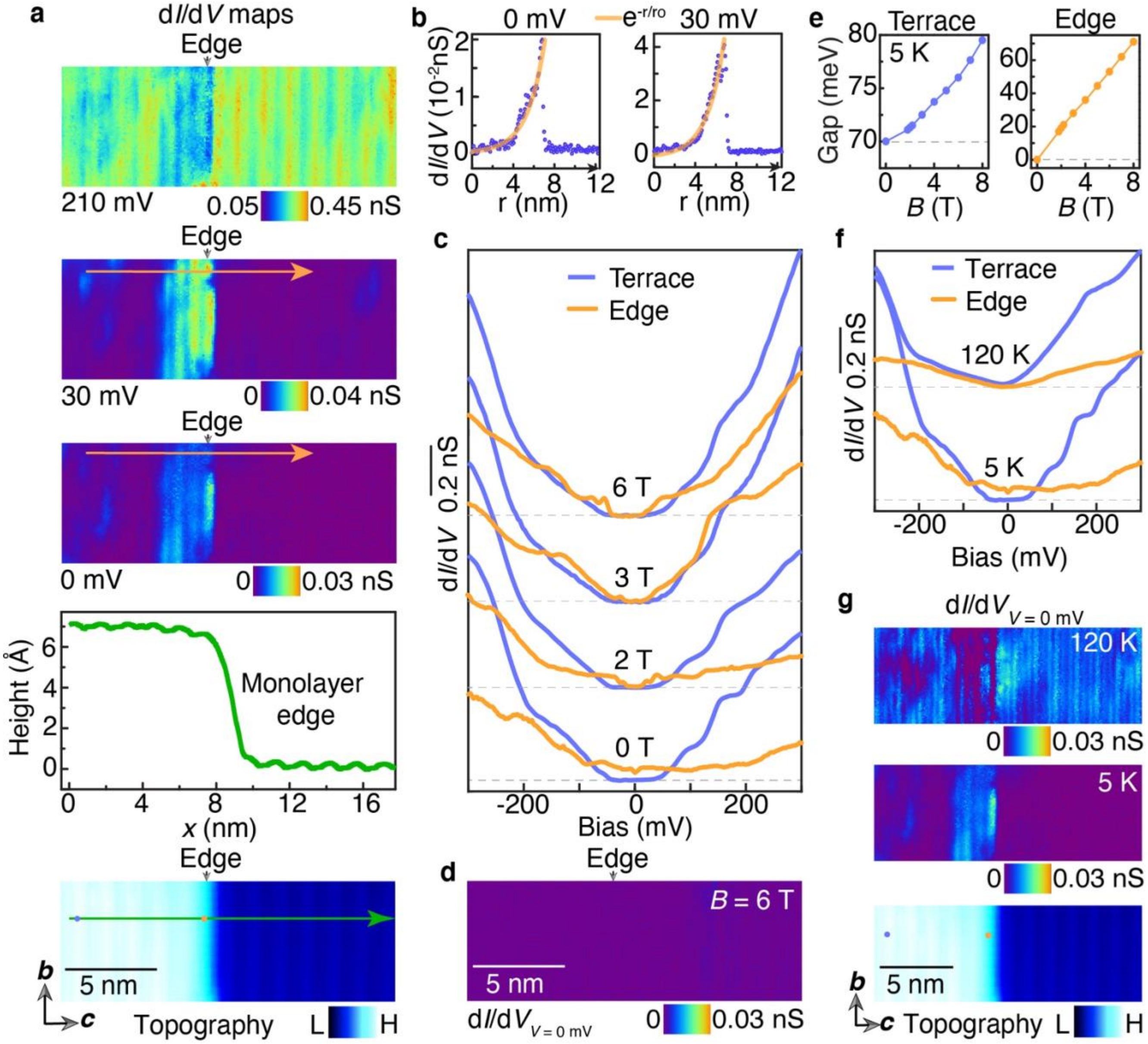

**Fig. 3: Topological nature of the insulating gap. a**, d*I*/d*V* maps acquired at different bias voltages (corresponding topography is shown in the bottom panel) around a monolayer step edge parallel to the *b*-axis, measured at $T = 5$ K. The height profile taken perpendicularly to the *b*-axis is also displayed. The d*I*/d*V* maps obtained within the energy gap ($V = 0$ mV and 30 mV) reveal a pronounced edge state, whereas, at $V = 210$ mV, the edge state is suppressed. **b**, Intensity distribution of the differential conductance around the step edge captured at $V = 0$ mV and 30 mV. The corresponding location is marked by a green line on the topographic image in panel **a,** and the scan direction is indicated by the arrow. The red curve represents the exponential fitting ($e^{-r/r_0}$) of the decay of the state away from the step edge, with a fitted decay length ($r_0$) of 1.6 nm. **c**, Tunneling spectra at locations away from the step edge and on the step edge measured under various magnetic fields. The orange and violet curves represent the differential spectra taken at the step edge and away from it, respectively. The corresponding spatial locations

where the spectra are acquired are indicated by the color-coded dots on the topographic image in panel **a.** Spectra at different magnetic fields are taken at the same locations and are vertically offset for clarity. Dashed horizontal lines mark the zero d$I$/d$V$ for different fields. At $B = 0$, tunneling spectra reveal an energy gap around the Fermi energy away from the step edge, while a pronounced gapless, in-gap state is observed on the step edge. In contrast, at $B$ = 2 T, 3 T, and 6 T, the step edge state is suppressed, and an energy gap gradually develops upon increasing magnetic field. **d**, d$I$/d$V$ map at $B$ = 6 T ($V$ = 0 mV) taken in the same region as in panel **a**, exhibiting suppressed differential conductance along the edge. **e**, Measured spectroscopic energy gaps at the terrace (left) and on the step edge (right) plotted as a function of magnetic field. Both energy gaps increase reasonably linearly with the magnetic field. **f**, Tunneling spectra at locations away from the step edge and on the step edge measured at temperatures $T =$ 120 K and 5 K. The orange and violet curves represent the differential spectra taken at the step edge and away from it, respectively. Spectra at the two temperatures are taken at the same locations and are vertically offset for clarity. Dashed horizontal lines mark the zero d$I$/d$V$ for different temperatures. At $T = 120$ K, the energy gap at the terrace is nonexistent, and the spectrum at the step edge becomes suppressed (with respect to the d$I$/d$V$ spectrum at the terrace) near the Fermi energy, visualizing the disappearance of the edge state. **g**, d$I$/d$V$ maps at the Fermi energy ($V = 0$ mV) around a monolayer step edge parallel to the $b$-axis measured at $T = 120$ K and 5 K. The corresponding topography is shown in the bottom panel. The color-coded dots overlaid on the topographic image correspond to the specific spatial locations where the d$I$/d$V$ spectra in panel **f** were acquired. The d$I$/d$V$ map at $T =$ 5 K shows a pronounced edge state, whereas at $T = 120$ K, the edge state is suppressed, indicating a direct correspondence between the emergence of the gapped bulk excitonic insulator state and the gapless edge state. Tunneling junction set-up: $V_{set}$ = 300 mV, $I_{set}$ = 0.5 nA, $V_{mod}$ = 2 mV.

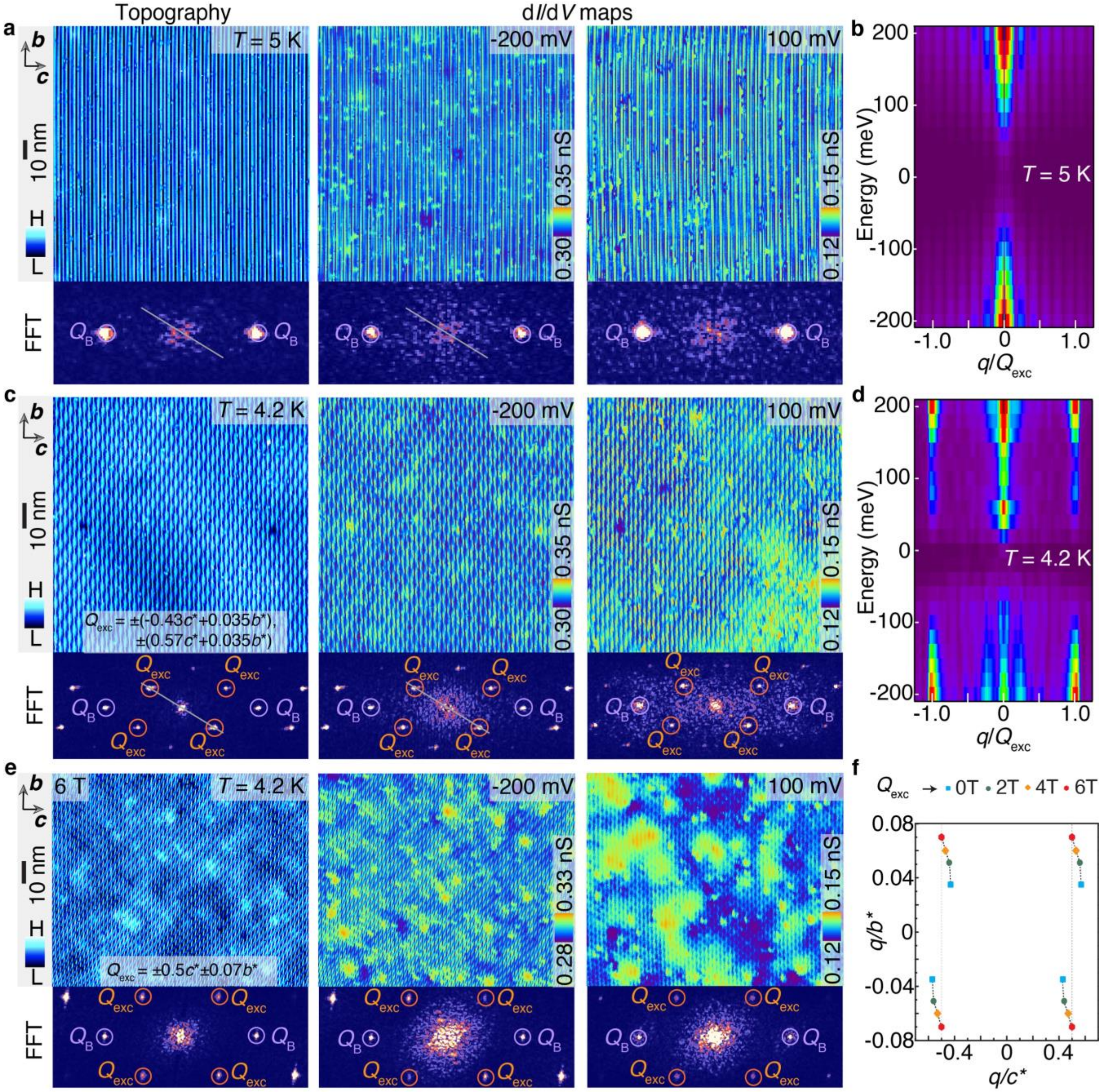

**Fig. 4: Observation of a secondary exciton instability with non-zero wavevector leading to translational symmetry breaking.** **a**, Topography (left panel) and d$I$/d$V$ maps acquired at the occupied ($V = -200$ mV) and unoccupied side ($V = 100$ mV) of the insulating gap measured at $T = 5$ K. The bottom panels show the corresponding Fourier transform images revealing well-developed Bragg peaks (purple circles). **b**, Energy dependence of the Fourier transform magnitude along the line cut marked by a grey line in the Fourier transform images (the horizontal axis is normalized by $Q_{\mathrm{exc}}$, defined in panel **c**). **c**, Topography (left panel) and d$I$/d$V$ maps acquired at the occupied ($V = -200$ mV) and unoccupied side ($V = 100$ mV) of the insulating gap, measured at $T = 4.2$ K, showing a pronounced translational symmetry breaking order as revealed through topographic and

spectroscopic contrast. Notably, there is a spectroscopic contrast reversal when the bias is switched between the occupied and unoccupied sides of the insulating gap. The Fourier transform images of the topography and the $\mathrm{d}I/\mathrm{d}V$ maps shown on the bottom panels display well-defined wavevector peaks (orange circles) alongside the Bragg peaks (purple circles). The wavevector, $Q_{\mathrm{exc}} = \pm(-0.43c^* + 0.035b^*), \pm(0.57c^* + 0.035b^*)$. **d**, Energy-dependent Fourier transform magnitude along the linecut indicated with a grey line in the Fourier transform images showing the energy dependence of $Q_{\mathrm{exc}}$; the linecut position is the same as in panel **a**. Akin to the data in panel **b**, the horizontal axis here is also normalized by $Q_{\mathrm{exc}}$. In contrast to the panel **b** data, however, a non-dispersing, well-developed peak intensity is seen at $q = Q_{\mathrm{exc}}$ and at both sides of the insulating energy gap. Plots in panels **b** and **d** were constructed using voltage bias intervals of 20 meV and 12.5 meV, respectively. As such, these plots lack sufficient energy resolution to accurately reflect gap sizes at the respective temperatures. Energy gaps are more reliably captured in the higher-resolution data displayed in Fig. 1**d**, **e**, where the bias interval is 1 meV. **e**, Topography (left panel) and $\mathrm{d}I/\mathrm{d}V$ maps acquired at the occupied ($V = -200$ mV) and unoccupied side ($V = 100$ mV) of the energy gap, measured at $T = 4.2$ K under an external magnetic field of 6 T, revealing the topographic and spectroscopic contrast associated to the translational symmetry breaking order. Akin to panel **c**, there is a spectroscopic contrast reversal when the bias is switched between the occupied and unoccupied sides of the insulating gap. The Fourier transform images of the topography and the $\mathrm{d}I/\mathrm{d}V$ maps shown on the bottom panels display well-defined wavevector peaks (orange circles) alongside the Bragg peaks (purple circles). Strikingly, here $Q_{\mathrm{exc}} = \pm(0.5c^* + 0.07b^*)$ is different from the $Q_{\mathrm{exc}}$ under zero magnetic field. **f**, Scatter plot highlighting the magnetic field tunability of $Q_{\mathrm{exc}}$. As the field increases, the wavevector $Q_{\mathrm{exc}}$ evolves gradually from being incommensurate along both the $b$- and $c$-axes at $B = 0$ T, to become commensurate along the $c$-axis under $B = 6$ T. Tunneling junction set-up: $V_{\mathrm{set}} = 300$ mV, $I_{\mathrm{set}} = 0.5$ nA, $V_{\mathrm{mod}} = 2$ mV.

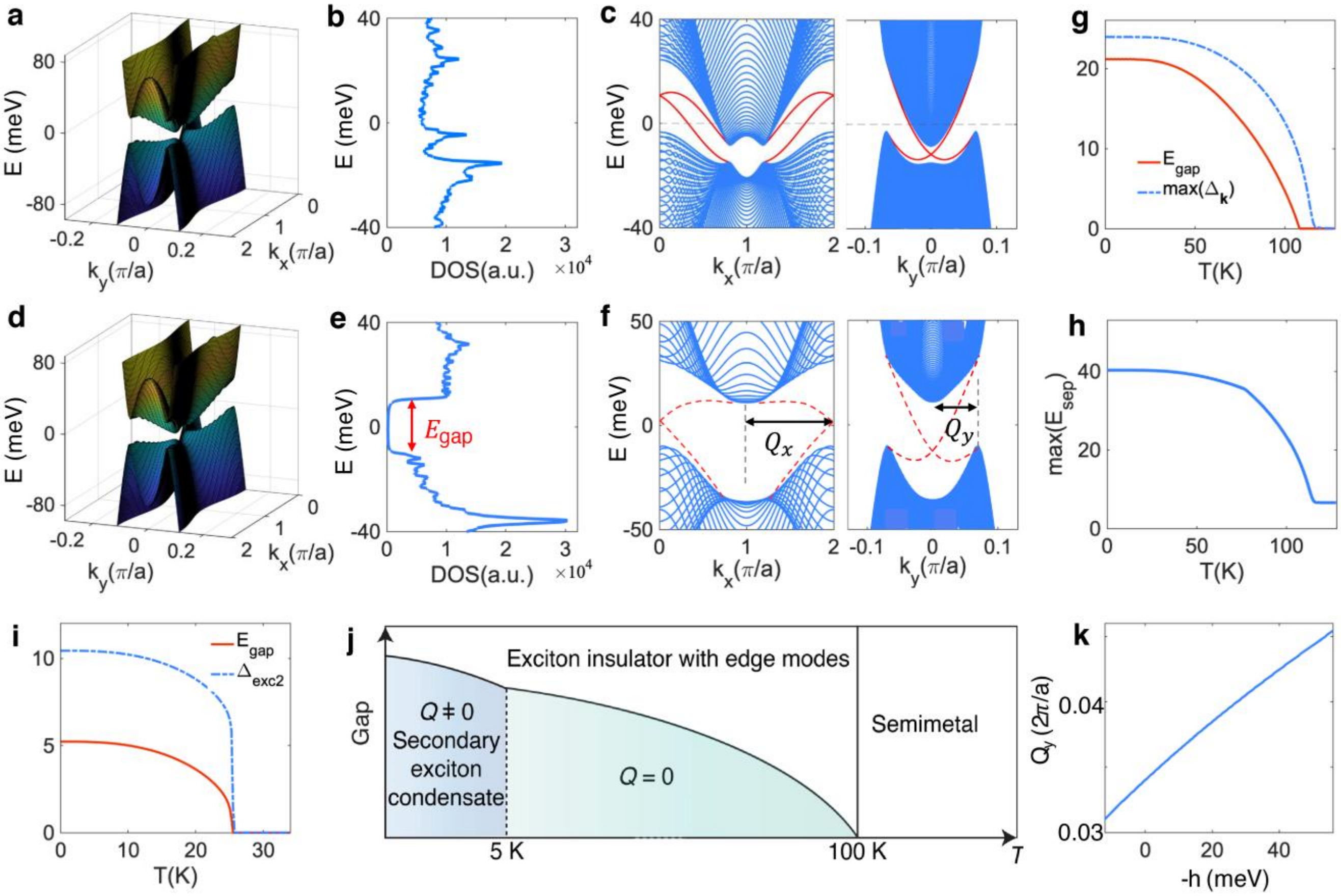

**Fig. 5: Theoretical modelling of the primary and secondary exciton condensation transition.** **a**, Low-energy band structure in the absence of an excitonic phase. **b**, Density of states in the absence of the excitonic phase. The system is metallic with the conduction band minimum located at 5.5 meV below the valence band maximum. **c**, Band structure for a ribbon with open boundaries along the *y*- (left) and *x*-directions (right), respectively. Blue bands represent the bulk continuum. Red curves illustrate the $Z_2$ topological edge states that connect the conduction and valence bands. **d**, Band structure in the presence of a primary excitonic order at $T = 10^{-3}$ K and $V = 1.16$ eV. **e**, Density of states in the presence of the primary excitonic order, exhibiting the opening of a fully developed bulk gap $E_{\text{gap}} \simeq 21.5$ meV. **f**, Projection of the band structure along the *x*- (left) and *y*- (right) directions. Red dashed curves indicate the edge states. $Q_x$ and $Q_y$ denote the momentum separation between the conduction band minimum and valence band maximum along the $k_x$- and $k_y$-directions, respectively. **g**, Band energy gap $E_{gap}$ (red curve) as a function of the $T$, exhibiting a monotonic decrease with increasing $T$ and an its disappearance for $T \geq 110\ K$, reminiscent of our experimental observation. Blue dashed curve plots the maximum value of the primary excitonic order parameter in momentum space max ($\Delta_{\mathbf{k}}$) as a function of $T$. **h**, Minimum separation between the conduction and valence bands min ($E_{sep}$) (red curve) as a function of $T$. The separation is always finite when changing the temperature. **i**, Secondary, finite momentum excitonic order parameter $\Delta_{\mathbf{exc2}}$ (blue dashed curve) and bulk energy gap $E_{gap}$ (red curve) as functions of $T$, both developing below 25 K. We consider the same Coulomb interaction strength as that used for the primary excitons. The secondary excitonic order is significantly weaker than the primary one and exhibits a lower critical temperature. **j**, Schematic phase diagram of the discovered phases of $Ta_2Pd_3Te_5$. **k**, Momentum separation, $Q_y$, between the valence band maximum and the $\Gamma$-point as a function of the Zeeman-like energy $h$.

# Methods:

### I. Single-crystal synthesis

$Ta_2Pd_3Te_5$ single crystals were synthesized using a flux method. A mixture of Ta powder (99.995% purity), Pd powder (99.95% purity), and Te chunks (99.99% purity) in a molar ratio of 2:4.5:7.5 was sealed in an evacuated quartz tube. All the manipulations were performed in an argon-filled glovebox to maintain an inert atmosphere. The sealed ampule was then heated to 950°C over a period of 10 hours and held at that temperature for 48 hours to allow for crystal growth. Subsequently, the ampule was gradually cooled to 800°C with a ramp rate of 2 K/h. At this temperature, any excess liquid was removed through a centrifuging process. The resulting $Ta_2Pd_3Te_5$ single crystals obtained from the growth process were shiny, rod-like, and exhibit a silver-gray color.

### II. Scanning tunneling microscopy

Single crystals were cleaved mechanically in situ at $T = 77$ K under ultra-high vacuum conditions ($< 5 \times 10^{-10}$ mbar). Immediately after cleaving, the crystals were inserted into the microscope head, which was already at the $^4$He base temperature (4.2 K). More than 20 single crystals were cleaved for this study. For each cleaved crystal, we explored surface areas over 10 μm × 10 μm to search for atomic flat surfaces. Topographic images were acquired in constant current mode. Tunneling conductance spectra were obtained with a commercial Ir/Pt tip (annealed under ultra-high vacuum condition and then characterized with a reference sample) using standard lock-in amplifier techniques with a lock-in frequency of 977 Hz. The tunneling junction set-ups used in the experiments are indicated in the corresponding figure captions. The magnetic field was applied through a zero-field cooling method. To acquire field-dependent tunneling conductance measurements, the tip was first withdrawn away from the sample, and then the magnetic field was slowly ramped to the desired value. The tip was then reapproached to the sample, and spectroscopic measurements were performed at the specific magnetic field. For temperature-dependent measurements, the tip was retracted from the sample, and the temperature was raised and stabilized for 12 hours before reapproaching the tip to the sample for tunneling measurements.

### III. Scanning transmission electron microscopy

Thin lamellae were prepared by focused ion beam cutting. All samples for the experiments were polished using a 2-kV gallium ion beam to minimize damage to the surface caused by the high-energy ion beam. Transmission electron microscopy (TEM) imaging, atomic-resolution high-angle annular darkfield scanning transmission electron microscopy (HAADF-STEM) imaging, and atomic-level energy-dispersive X-ray spectroscopy (EDS) mapping were performed using a Titan Cubed Themis 300 double Cs-corrected scanning/transmission electron microscope. The microscope was equipped with an extreme field emission gun source and operated at 300 kV. Additionally, a super-X energy-dispersive spectrometry system was utilized during the experiments.

### IV. X-ray diffraction measurements

Single crystal X-ray diffraction measurements were performed using a custom-designed X-ray diffractometer equipped with a Xenocs Genix3D Mo Kα (17.48 keV) X-ray source, delivering $2.5 \times 10^7$ photons/sec in a beam spot size of 150 μm at the sample position. The samples under investigation were mounted on a Huber 4-circle diffractometer and cooled using a closed-cycle cryostat with a beryllium dome. The diffraction signals were captured by a highly sensitive single photon counting PILATUS3 R 1M solid-state area detector, boasting 981 × 1043 pixels, each with a size of 172 μm ×172 μm. During data acquisition, images were taken in 0.1° increments while rotating the samples, and these images were subsequently converted into three-dimensional mappings in momentum space. The crystallographic structure of the samples was confirmed to conform to the space group *Pnma*.

The lattice parameters at room temperature were determined to be $a = 13.973$ Å, $b = 3.7098$ Å, and $c = 18.564$ Å, with a measurement system error of 0.3%.

## V. Angle-resolved photoemission spectroscopy

Synchrotron-based angle-resolved photoemission spectroscopy (ARPES) measurements were performed on single crystals, cleaved in ultra-high vacuum ($< 1 \times 10^{-10}$mbar) on the ARPES manipulator. High-resolution energy-momentum data shown in Fig. 2 and Supplementary Figs. 1-4 were collected at the beamline 5-2 of the of the Stanford Synchrotron Radiation Lightsource (SSRL) in Stanford, USA. To avoid the nonreversible spectrum broadening due to degradation, all the temperature-dependent measurements were performed by cooling down a sample that was freshly cleaved at $T = 350$ K. The energy and momentum resolution were better than 10 meV and 0.008 $\text{Å}^{-1}$. The preliminary data were taken at the beamline 4 of the Advanced Light Source (ALS) in Berkeley, USA, and National Synchrotron Light Source in Brookhaven, USA.

## VI. First-principles calculations

Electronic band structure calculations were performed within the density functional theory framework using the Vienna Ab initio Simulation Package (VASP)[60, 61]. The General gradient approximation functional is used for treating the exchange-correlation effect[62]. The relativistic effect of spin-orbit coupling was included self consistently in the calculations. An energy tolerance of $10^{-6}$ eV was used. We performed the calculations using 4×8×4 k-mesh centered at the Γ-point for Bulk, and 1×8×4 k-mesh for monolayer. We extracted the real space tight-binding Hamiltonian by Wannier function using the Wannier90 code[63] with *d*-orbitals of Ta and Pd, and *p*-orbitals of Te as projections. The Wannier tight binding Hamiltonian was used to study the topological properties using the Wannier tools package[64].

## VII. Electrical transport measurements

To fabricate the $Ta_2Pd_3Te_5$ sample for transport measurements, we employed a polydimethylsiloxane (PDMS) stamp-based mechanical exfoliation technique. First, we patterned the sample contacts on the Si substrates with a 280 nm layer of thermal oxide using electron beam lithography, followed by chemical development and metal deposition (5 nm Cr/35 nm Au). The fresh $Ta_2Pd_3Te_5$ flakes were mechanically exfoliated from bulk single crystals on PDMS stamps. We carefully selected thick flakes (to capture the bulk properties of $Ta_2Pd_3Te_5$) with good geometry through optical microscopy. Next, we transferred the flakes onto the $SiO_2$/Si substrates that had pre-patterned Cr/Au electrodes in place. To preserve the intrinsic properties of the compound and minimize environmental effects, we encapsulated the samples using thin polymethyl methacrylate films with thicknesses around ~50 nm, which ensured that the samples on the devices were never exposed to air directly. All sample fabrication processes were performed in a glovebox equipped with a gas purification system (<1 ppm of $O_2$ and $H_2O$).

Electrical transport measurements were conducted using an Oxford Heliox system with a base temperature of 0.3 K and a maximum magnetic field up to 8 T. We employed a standard four-probe method to measure the resistance and differential resistance of the sample, utilizing a lock-in amplifier technique with a lock-in frequency of 13 Hz. The reproducibility of the results was confirmed by conducting measurements in the National High Magnetic Field Laboratory in Tallahassee, Florida, USA. To ensure consistency, multiple devices were prepared and measured over multiple runs.

## VIII. No detection of a structural instability in $Ta_2Pd_3Te_5$

As the presence of a structural instability can complicate the experimental detection of the transition to the excitonic insulator state, it is crucial to investigate whether the observed excitonic insulator transition in $Ta_2Pd_3Te_5$ is accompanied by any structural instabilities. For this purpose, we conducted a detailed structural analysis via transmission electron microscopy. Our scanning transmission electron microscopy data reveals consistent atom arrangements compared to those of the pristine $Ta_2Pd_3Te_5$ crystal structure on different surfaces (Extended Fig. 1**a**,**c** and Extended Fig. 2). Importantly, the selected area electron diffraction patterns obtained from the focused-ion-beam cut lamella at $T = 290$ K and 90 K exhibit identical crystal lattice at both temperatures (Extended Fig. 1**b**,**d**). This observation suggests that the potential structural change associated to the excitonic transition at $T = 100$ K is below our detection threshold.

We have conducted X-ray diffraction measurements at different temperatures (ranging from $T = 300$ K to 20 K) to strengthen our conclusion regarding the structural changes in $Ta_2Pd_3Te_5$. The results are summarized in Extended Fig. 3. The data closely matches the expected *Pnma* structure across the entire temperature range. The potential t-subgroups of space group # 62 (*Pnma*) that are compatible with the mirror symmetry breaking in the $q = 0$ excitonic insulator phase detected via our photoemission measurements are subgroups 19 ($P2_12_12_1$), 4 ($P2_1$), 2 ($P$-1), and 1 ($P1$). However, determining the exact subgroup will require future ultra-high-resolution structural measurements. We refer the reader to Supplementary Notes 8 and 9 and Supplementary Figs. 9 and 10 for additional data and discussion.

## IX. Energy gap determination from the tunneling spectra

In this section, we present a procedure that we utilize to determine the value of the spectroscopic energy gap from $dI/dV$ curves. The key idea is to identify the edges of the energy gap, where a discernible non-zero $dI/dV$ signal can be distinguished from the noisy zero $dI/dV$ signal within the gap region[65]. The following steps outline our approach to determining the energy gap from a single $dI/dV$ curve (refer to Extended Fig. 4):

1. Initially, we select a bias range where the $dI/dV$ signal takes the form: $dI/dV = 0 + \xi$, where 0 represents the mean $dI/dV$ value within the spectroscopic gap and $\xi$ represents the noise in the $dI/dV$ signal. This is done for a bias voltage range $V \in [V_1, V_2]$, while avoiding the edges of the spectroscopic gap.
2. Next, we calculate the standard deviation of the noisy $dI/dV$ signal within the gap, denoted as $\sigma = \sqrt{\overline{\xi^2}}$. By determining $\sigma$ from the gap segment of the $dI/dV$ curve, we establish a noise floor for the $dI/dV$ signal.
3. We assume a Gaussian distribution of $\xi$ and define $\Gamma = 2.36\sigma$, which represents the full width at half maximum. Two Gaussian signals can be distinguished if the difference between the means of the signals is greater than the full width at half maximum. Thus, we set $\Gamma$ as the instrumental resolution of the $dI/dV$ signal.
4. A non-zero $dI/dV$ signal can only be detected when $dI/dV > \Gamma$ within our instrumental resolution. We set $\Gamma$ as the threshold value for identifying the edges of the energy gap.
5. By solving the equation $dI/dV = \Gamma$, we determine the intersections $V_a$, $V_b$ of the threshold value with the $dI/dV$ spectroscopic curve. $eV_a$ and $eV_b$ represent the energies of the gap edge above and below $E_F$, respectively. Finally, we calculate $\Delta = eV_a - eV_b$ to obtain the value of the spectroscopic energy gap.

## X. Broken symmetries due to excitonic insulator transition

The excitons form due to the attractive Coulomb interaction between electron and hole states. This phenomenon is an electronic property related to the spontaneous breaking of particle-hole U(1) symmetry, akin to BCS superconductivity. In addition to U(1) symmetry, the excitonic order may further break inversion and mirror symmetries.

In the absence of excitonic order, the conduction and valence bands at the Γ-point have different parity, meaning that they are eigenstates of inversion symmetry with different eigenvalues. The excitonic order hybridizes the conduction and valence bands, thereby violating this symmetry.

To elucidate this, we examined the symmetries of our model (see Methods Section XIV for the details of our model) both with and without the $Q = 0$ excitonic order. In the absence of excitonic order, the effective Hamiltonian is given by:

$$H(k_x, k_y) = h(k_x, k_y)\sigma_0\, s_0 + \mathcal{M}(k_x, k_y)\sigma_z\, s_0 + A_x sink_x \sigma_x\, s_z + A_y sink_y\, \sigma_y\, s_0, \quad (1)$$

where $h(k_x, k_y)$ and $\mathcal{M}(k_x, k_y)$ are even functions of $k_x$ and $k_y$. This model exhibits inversion and mirror symmetries, as indicated by:

$$PH(k_x, k_y)P^{-1} = H(-k_x, -k_y), \quad (2a)$$

$$M_x H(k_x, k_y) M_x^{-1} = H(k_x, -k_y), \quad (2b)$$

$$M_y H(k_x, k_y) M_y^{-1} = H(-k_x, k_y), \quad (2c)$$

where $P = \sigma_z$ , $M_x = i\sigma_z\, s_x$ and $M_x = i\sigma_0\, s_y$ are the inversion and mirror symmetry operators, respectively.

Applying the unitary transformation, we can diagonalize the Hamiltonian as:

$$H' = UH(k_x, k_y)U^{-1} = E_c(k_x, k_y)(\sigma_0 + \sigma_z)s_0/2 + E_v(k_x, k_y)(\sigma_0 - \sigma_z)s_0/2, \quad (3)$$

where $E_c(k_x, k_y)$ and $E_v(k_x, k_y)$ are the conduction and valence bands. The transformation matrix is given by $U = U_1 \oplus U_2$, with

$$U_1 = (cos(\phi/2), sin(\phi/2)e^{i\theta};\ -sin(\phi/2)e^{-i\theta}, cos(\phi/2)), \quad (4a)$$

$$U_2 = (cos(\phi/2), -sin(\phi/2)e^{-i\theta};\ sin(\phi/2)e^{i\theta}, cos(\phi/2)), \quad (4b)$$

where $\phi$ and $\theta$ are determined by $\tan\phi = \sqrt{(A_x sink_x)^2 + (A_y sink_y)^2}/\mathcal{M}(k_x, k_y)$ and $e^{i\theta} = (A_x sink_x + iA_y sink_y)/\sqrt{(A_x sink_x)^2 + (A_y sink_y)^2}$.

Now, we consider the excitonic order. In the band basis, the exciton order parameter takes the form:

$$H'_{exc} = H' + \Delta\,(k_x, k_y)\boldsymbol{\sigma}_x s_0. \quad (5)$$

Transforming back to the spin and orbital basis, the full Hamiltonian $H_{exc} = U^{-1}H'_{exc}U$ is given by

$$H_{exc}(k_x, k_y) = H(k_x, k_y) + \Delta\,(k_x, k_y)M, \quad (6)$$

where $M = M_1 \oplus M_2$ with

$$M_1 = (-cos\theta sin\phi, e^{i\theta}(cos\theta\ cos\phi - i\ sin\theta);\ e^{-i\theta}(cos\theta\ cos\phi + i\ sin\theta), cos\theta sin\phi), \quad (7a)$$

$$M_2 = (cos\theta sin\phi, e^{-i\theta}(cos\theta\ cos\phi + i\ sin\theta);\ e^{i\theta}(cos\theta\ cos\phi - i\ sin\theta), -cos\theta sin\phi). \quad (7b)$$

At the Gamma point, $M_1 = M_2 = (0{,}1;\ 1{,}0)$. We see that $H_{exc}(k_x, k_y)$ no longer obeys the symmetry relations in Eqs. 2(a)-(c) due to the presence of the exciton term $\Delta\,(k_x, k_y)M$. This indicates that the inversion and mirror symmetries are explicitly broken.

## XI. Extended scanning tunneling microscopy on the edge states

The existence of edge states in $Ta_2Pd_3Te_5$ was reported in a previous work[31]. In our study, presented in Fig. 3, we took an important step beyond the observation of the edge state by providing compelling evidence for its exponential localization and time-reversal-symmetry-protected, $Z_2$ topological nature. In this section, we further support the data presented in Fig. 3 by providing extended results on the time-reversal-symmetry-protected nature of the edge state along different edge configurations. Extended Fig. 5 summarizes our STM measurements on two types of atomic step edges: (i) a monolayer step edge running along the *c*-axis, perpendicular to the one-dimensional Te chains, and (ii) a four-atomic-layer thick step edge along the *b*-axis.

In Extended Fig. 5**a**, we show a monolayer atomic step edge along the *c*-axis, identified through the topographic image and corresponding height profile. Notably, the spatially resolved d*I*/d*V* map at the Fermi energy, within the insulating bulk gap, reveals a pronounced edge state, akin to the step edge along the *b*-axis shown in Fig. 3**a**. Energy-resolved d*I*/d*V* spectra on the step edge (Extended Fig. 5**b**) exhibit a substantial d*I*/d*V* signal (orange curves) around the Fermi energy, while the d*I*/d*V* spectrum away from the step edge (violet curves) displays an insulating gap. Additionally, akin to the monolayer step edge along the *b*-axis, the application of an external magnetic field perpendicular to the *bc*-plane leads to significant suppression of the d*I*/d*V* measured at the step edge. The field-dependent tunneling spectra, shown for $B = 0$ T, 2 T, and 4 T in Extended Fig. 5**b**, demonstrate a gradual opening of an insulating gap at the edge state spectrum, confirming its time-reversal symmetry protected nature. The presence of the edge state and its time-reversal symmetry protected nature is also observed in multilayer step edges. For instance, we present topography, corresponding height profile, and spectroscopic measurements on a four-atomic-layer thick step edge in Extended Figs. 5**c** and 5**d**. The spectroscopic mapping reveals a pronounced gapless in-gap state at the Fermi energy at $B = 0$, while the tunneling spectra at $B = 0$ T, 2 T, and 4 T exhibit the progressive opening of an energy gap with an increasing magnetic field, highlighting the presence of a time-reversal-symmetry-protected, topological edge state in the four-layer step edge.

## XII. Extended scanning tunneling microscopy data on the field tunability of the secondary translational symmetry breaking excitonic order

Here, we present high-resolution topographic images obtained through STM, revealing the translational symmetry breaking ordering in $Ta_2Pd_3Te_5$. In Extended Fig. 6**a**, we display an atomically resolved topographic image of a pristine $Ta_2Pd_3Te_5$ (100) surface acquired at $T = 5$ K. The inset shows the corresponding Fourier transform image, indicating Bragg peaks along the $c$-axis. It is worth noting that we are focusing on a small region in the Fourier transform image, and the Bragg peaks along the $b$-axis are located further away in $q$-space (refer to Fig. 1). At $T = 5$ K, no wavevector peaks are visible. However, when the sample is cooled to $T = 4.2$ K, within the same region as in Extended Fig. 6**a**, we observe clear translational symmetry breaking in real space (Extended Fig. 6**b**). The corresponding Fourier transform image, plotted in the same range as the inset in Extended Fig. 6**a**, now exhibits distinct superlattice peaks labeled as $Q_{\mathrm{exc}}$.

Having observed the emergence of the translational symmetry breaking superlattice order at $T = 4.2$ K, we proceed to investigate its magnetic field dependence. In Extended Fig. 6**c**-**e**, we present topographic images captured under magnetic fields of 2 T, 4 T, and 6 T, respectively. The topographic images clearly demonstrate the evolution of the superlattice modulation pattern as the magnetic field changes. This evolution is further highlighted in the corresponding Fourier transform images, all plotted within the same $q$ range, which illustrate the variation of $Q_{\mathrm{exc}}$ with increasing magnetic field. Notably, the positions of the Bragg peaks ($Q_{\mathrm{B}}$) remain unaffected by the magnetic field, as expected. To quantitatively summarize the change in $Q_{\mathrm{exc}}$ as a function of the magnetic field, Extended Fig. 6**f** displays a polar plot of the vector, $Q_{\mathrm{exc}}$ at various magnetic field strengths (a scatter plot for $Q_{\mathrm{exc}}$ at different magnetic fields is shown in Fig. 4**f**). The plot clearly demonstrates the gradual evolution of $Q_{\mathrm{exc}}$ with increasing magnetic field, in accordance with the behavior expected for a finite momentum exciton condensate. This behavior may arise due to the Zeeman effect experienced by the particle and hole bands associated with the exciton condensate upon the application of a magnetic field. The resulting Zeeman shift in the bands causes a change in the exciton wavevector, which in real space, leads to the variation of $Q_{\mathrm{exc}}$.

## XIII. Transport evidence of the translational symmetry breaking secondary excitonic order and its field tunability

In this section, we present electrical transport evidence to support the formation of a translational symmetry breaking order, indicating that it is a bulk phenomenon rather than a surface effect.

To conduct systematic electrical transport measurements, we fabricated four-point probe devices (as shown in Extended Fig. 7**a**) using mechanically exfoliated flakes that are thick enough to capture the bulk properties of $Ta_2Pd_3Te_5$. The sample resistance as a function of the temperature is displayed in Extended Fig. 7**b**. The left inset of Extended Fig. 7**b** highlights a slight decrease in the resistance as a function of temperature until $T \simeq 100$ K is reached, indicating a semimetallic behavior. However, below $T \simeq 100$ K, the resistance starts to increase rapidly as the temperature decreases, exhibiting insulating behavior. Additionally, a small kink in the resistance is observed near $T \simeq 5$ K (right inset in Extended Fig. 7**b**). We refer the reader to Supplementary Fig. 12 for data from a second sample.

To investigate the translational symmetry breaking order, we explore the motion of this order under a DC electric field. The translational symmetry breaking order is expected to be pinned by the ubiquitous disorder potential created by impurities but can be de-pinned[66] under a sufficiently large electric field causing the electronic crystal to slide. Therefore, the current-voltage characteristics should exhibit non-linear behavior indicative of a sliding phason mode[66-68]. To explore this possibility, we applied a DC current between the source and drain electrodes and measured the sample's differential resistance ($dV/dI$) using a standard lock-in technique. We also measured the DC voltage across the voltage leads and extracted the electric field using $E = V/L$ ($L$ is the distance between the voltage probes). The measured threshold voltage ($V_{th}$) is found to have a linear relationship with $L$ (top inset of Extended Fig. 7**d**), justifying the use of $E = V/L$ relationship. Extended Fig. 7**c** shows a plot of $dV/dI$ as a function of $E$ at various temperatures. At small values of $E$, $dV/dI$ remains constant for all temperatures. However, below $T = 5.1$ K, $dV/dI$ sharply decreases above a threshold electric field ($E_{th}$), indicating the onset of a collective phason current[58]. The temperature dependence of $E_{th}$, defined as the onset of the sharp decrease in $dV/dI$ with respect to $E$, is illustrated in Extended Fig. 7**d**. We find that $E_{th}$ disappears above $T = 5.1$ K, which we identify as the transition temperature ($T_{exc2}$) of the translational symmetry breaking order.

Additionally, it is crucial to consider the impact of the Joule heating on the observed transport properties. If heat were dissipated within the sample, leading to a localized temperature increase above $T_{exc2}$, it could potentially induce a thermally driven switching effect and result in a non-linear transport characteristics. However, in such a scenario, the required dissipated power for switching would tend to approach zero as $T$ approaches $T_{exc2}$ from below. Contrary to this expectation, the electrical power $P_{th} = V_{th}^2/[dV/dI(V_{th})]$ obtained at $V_{th}$ from our experiments does not approach zero as $T$ approaches $T_{exc2}$ (bottom inset of Extended Fig. 7**d**). Instead, $P_{th}$ decreases with decreasing temperature, which is precisely the opposite of what we would anticipate in a thermally driven switching scenario. Furthermore, it is important to note that the power dissipated in the sample, which is less than 1 pW, is significantly smaller than the cooling power of the cryostat. Moreover, the observed linear dependence of $V_{th}$ on the distance between the voltage leads (top inset of Extended Fig. 7**d**) suggests that the observed characteristics are not attributed to contact effects. Instead, these observations provide compelling evidence for the presence of a propagating, current-carrying translational symmetry breaking order.

Having discussed the transport evidence for the translational symmetry breaking order, we proceed to investigate how its properties evolve as a function of the magnetic field. In Extended Fig. 7**e**, we present $dV/dI$ as a function of $E$ at various magnetic fields, applied perpendicularly to the sample plane, ranging from -3 T to 7 T. The plot clearly exhibits a change in $E_{th}$ as a function of the magnetic field. This change is further quantified in Extended Fig. 7**f**, which demonstrates an increase in $E_{th}$ with an increasing magnetic field. The magnetic field-dependent behavior of the translational symmetry breaking order aligns with our STM observations (Fig. 4 and Extended Fig. 6), where we observe a gradual change in $Q_{exc}$ with increasing field. Notably, at $B = 0$, the translational symmetry breaking order is incommensurate and therefore not pinned to the lattice, explaining the small $E_{th}$ required to de-pin the translational symmetry breaking order from the pinning potential solely contributed by the impurities. Conversely, at $B = 6$ T, the translational symmetry breaking order becomes commensurate along one axis in the cleaving plane (the *c*-axis), potentially making it more difficult to de-pin the translational symmetry breaking order and requiring a larger $E_{th}$. This is consistent with our observations in Extended Fig. 7**f**. It is important to note that even though the translational symmetry breaking order becomes commensurate along the *c*-axis at $B = 6$ T, it may or may not be directly pinned to the lattice along that direction. Overall, our electrical transport experiments, demonstrating a well-defined $E_{th}$ that is consistent with the presence of the translational symmetry breaking order. In addition, the agreement of its onset temperature with the STM observations, serves as further bulk evidence for the existence of a bulk translational symmetry breaking order in $Ta_2Pd_3Te_5$ below $T \simeq 5$ K.

## XIV. Theoretical description of the two excitonic instabilities in $Ta_2Pd_3Te_5$

In our experiments, the edge states are gapless in the absence of magnetic fields. This observation suggests the presence of time-reversal symmetry, which prevents the coupling between the two spin species. This allows us to address the two spin species separately. Therefore, we focus on one spin species to elucidate the essential physics, while the results for the other spin species can be straightforwardly obtained by exploiting the time-reversal symmetry. The low-energy band structure for monolayer $Ta_2Pd_3Te_5$, derived from first-principles calculations, is displayed in Extended Fig. 8. The conduction band minimum is situated at $\boldsymbol{k} = \left(0, \frac{\pi}{c}\right)$, while the valence band maxima are located close to the $k_y = 0$ line. Notably, the system is gapless in the absence of the excitonic order. Specifically, the bottom of the conduction band minimum is situated slightly lower (by 6 meV) with respect to the top of the valence band maxima. Moreover, the band structure is topological and can be characterized by a nonzero topological invariant. Based on these properties, we constructed a two-band minimum model on a square lattice as follows:

$$H = \sum_{\boldsymbol{k}} \Psi_{\boldsymbol{k}}^{\dagger} \{\mathcal{M}(\boldsymbol{k})\sigma_z + A_x \sin k_x \sigma_x + A_y \sin\ k_y \sigma_y + h(\boldsymbol{k})\sigma_0\} \Psi_{\boldsymbol{k}}. \tag{8}$$

In Eq. (8), $\Psi_{\boldsymbol{k}} = \left(c_{\alpha,\boldsymbol{k}}, c_{\beta,\boldsymbol{k}}\right)^T$, $\mathcal{M}(\boldsymbol{k}) = M_0 + M_x(1 - \cos k_x) + M_y\left(1 - \cos k_y\right)$, and $h(\boldsymbol{k}) = C_x(1 + \cos k_x) + C_y(1 - \cos k_y)$. $\{\sigma_x, \sigma_y, \sigma_z\}$ and $\sigma_0$ are Pauli and identity matrices for pseudo-spin $\{\alpha, \beta\}$, and the lattice constants are set to be unity. It is essentially a modified Bernevig-Hughes-Zhang model[69].

The two energy bands are given by

$$\varepsilon_{c\,/v}(\boldsymbol{k}) = h(\boldsymbol{k}) \pm E(\boldsymbol{k}), \tag{9}$$

where $E(\boldsymbol{k}) = \left[\mathcal{M}(\boldsymbol{k})^2 + A_x^2 \sin^2 k_x\ + A_y^2 \sin^2 k_y\ \right]^{\frac{1}{2}}$. To mimic the low-energy band structure of monolayer $Ta_2Pd_3Te_5$, we chose the following parameters: $M_0 = -0.072$ eV, $M_y = 3.2$ eV, $M_x = 0.04$eV, $A_x = 0.006$ eV, $A_y = 0.064$ eV, $C_x = 0.003$ eV, and $C_y = -0.24$ eV. The resulting band structure is illustrated in Fig. 5**a**. The conduction band minimum is located slightly lower (by 5.5 meV) with respect to the valence band maxima. As dictated by the nontrivial topological invariant, the model exhibits gapless edge states under open boundaries. However, the edge states are buried by the bulk continuum (Fig. 5**c**).

**Topological excitonic instability:** We consider the Coulomb interaction, which in momentum space can be expressed as[70]:

$$H_{\text{int}} = \frac{1}{N} \sum_{\boldsymbol{k},\boldsymbol{k}',\boldsymbol{q}} W(\boldsymbol{q}) c_{\boldsymbol{k}+\boldsymbol{q},c}^{\dagger} c_{\boldsymbol{k}'-\boldsymbol{q},v}^{\dagger} c_{\boldsymbol{k}',v} c_{\boldsymbol{k},c}, \tag{10}$$

where $c_{\boldsymbol{k},c(v)}^{\dagger}$ and $c_{\boldsymbol{k},c(v)}$ represent creation and annihilation operators for an electron with momentum $\boldsymbol{k}$ at the conduction (valence) band, respectively; $N$ is the number of $\boldsymbol{k}$-points in the Brillouin zone; $W(\boldsymbol{q})$ stands for the $q$-resolved screened Coulomb interaction

$$W(\boldsymbol{q}) = \frac{U(\boldsymbol{q})}{1\ +\ 2\pi\alpha|\boldsymbol{q}|}, \tag{11}$$

and $U(\boldsymbol{q}) = \frac{V}{|\boldsymbol{q}|}$ denotes the unscreened Coulomb interaction which is the dominant contribution to the electron-hole attraction. We consider the two-dimensional screening effect which leads to an intrinsic $\boldsymbol{q}$-dependent dielectric function of the form $1/(1 + 2\pi\alpha|\boldsymbol{q}|)$, as shown in Eq. (11)[71, 72]. $V$ represents the interaction strength and $\alpha$ parametrizes the screening strength. In the Coulomb interaction described by Eq. (10), the summation involves three momenta, $\boldsymbol{k}$, $\boldsymbol{k}'$, and $\boldsymbol{q}$. To simplify this for the Hartree-Fock approximation, we impose specific constraints on the relationship between these momenta, guided by the electronic band structure of the material. Specifically, we consider $\boldsymbol{q} = \boldsymbol{k}' - \boldsymbol{k}$ and $\boldsymbol{q} = \boldsymbol{k}' - \boldsymbol{k} \pm \boldsymbol{Q}$ in Eq. (1), where $\boldsymbol{Q}$ is the ordering wavevector (corresponding to the distance between the positions of the conduction minimum and valence maximum). Consequently, the Coulomb interaction can be rewritten as two terms (components):

$$H_{\text{int}} = \frac{1}{N}\sum_{\boldsymbol{k},\boldsymbol{k}'} W(\boldsymbol{k}' - \boldsymbol{k}) c^{\dagger}_{\boldsymbol{k}',c} c^{\dagger}_{\boldsymbol{k},v} c_{\boldsymbol{k}',v} c_{\boldsymbol{k},c} + \frac{1}{N}\sum_{\boldsymbol{k},\boldsymbol{k}'} W(\boldsymbol{k}' - \boldsymbol{k} \pm \boldsymbol{Q}) c^{\dagger}_{\boldsymbol{k}'\mp\boldsymbol{Q},c} c^{\dagger}_{\boldsymbol{k}\pm\boldsymbol{Q},v} c_{\boldsymbol{k}',v} c_{\boldsymbol{k},c}. \quad (12)$$

Note that the two terms arise from different restrictions of $\boldsymbol{q}$ in Eq. (1), and, therefore, are independent. With these restrictions, the summation now involves only *two* momenta $\boldsymbol{k}$ and $\boldsymbol{k}'$, enabling us to perform the Hartree-Fock decomposition.

We first focus on the first term, $Q_y$, in Eq. (12) and decouple it as

$$H_{exc} = \begin{pmatrix} 0 & \Delta_{\boldsymbol{k}} \\ \Delta_{\boldsymbol{k}} & 0 \end{pmatrix} \quad (13)$$

in the band basis $(c_{\boldsymbol{k},c}, c_{\boldsymbol{k},v})$. The momentum-dependent order parameter $\Delta_{\boldsymbol{k}}$ is calculated self-consistently through

$$\Delta_{\boldsymbol{k}} = -\frac{1}{N}\sum_{\boldsymbol{k}'} W(\boldsymbol{k} - \boldsymbol{k}') \langle c^{\dagger}_{\boldsymbol{k}',c} c_{\boldsymbol{k}',v} \rangle, \quad (14)$$

where $\langle \ldots \rangle$ stands for the quantum statistic average. It yields the primary excitonic order that couples electrons and holes possessing identical momenta $\boldsymbol{k}$. Importantly, it should be emphasized that the order parameter is solely influenced by the Fock term. The influence of the Hartree term is detailed in Supplementary Note 14.

Through a self-consistent computation, utilizing an interaction strength of $V = 1.16$ eV, a small screening factor $\alpha = 0.3$ (in units of the lattice constant), and employing a 201 × 561 grid within the Brillouin zone, we observe the distinctive emergence of the excitonic order. This order engenders a band gap for temperatures $T < T_c \approx 110$ K. The order parameter $\Delta_{\boldsymbol{k}}$ is most pronounced near the $k_x$ axis (i.e., $k_y = 0$) where the two bands are closest to each other. With decreasing $T$, $\Delta_{\boldsymbol{k}}$ becomes stronger, in turn leading to a consistent increase in the bulk gap. Ultimately, for $T \leq 30$ K, the bulk gap saturates at a large value of 21.5 meV. Notably, the valence maxima relocate to $\pm(0, 0.071\pi)$ (as depicted in Fig. 5**d**).

In Fig. 5**g**, we calculate the maximum value of the order parameter, max ($\Delta_{\boldsymbol{k}}$), as a function of *T*. We find that the critical temperature for $\Delta_{\boldsymbol{k}}$ is $T_c' \approx 120$ K, which significantly exceeds $T_c$. This deviation is due to the negative band gap in the bared band structure (without $\Delta_{\boldsymbol{k}}$). Additionally, Fig. 5**h** portrays the minimum value of the energy separation between the conduction and valence bands, min ($E_{sep}$), as a function of *T*. Notably, $\Delta E_{min}$ remains consistently positive throughout the entire range of *T*s. This observation indicates that, as we turn on the excitonic order by decreasing temperature, the two bands still do not touch with each other. Thus, the topology of the system remains preserved.

We note that in refs.[73, 74], exciton condensation in the Bernevig-Hughes-Zhang model was studied under the assumption of local interaction. It was shown that the interaction spontaneously breaks time-reversal, spin $U(1)$, and parity symmetries, eventually driving the system into a trivial band insulator. However, this cannot explain our experiments, where we observe gapless edge modes that become more robust and localized within the exciton phase.

**Excitonic translational symmetry breaking order:** Next, we consider the second term in the Coulomb interaction Eq. (12), which involves the ordering wave wavevector $\boldsymbol{Q} \approx (\pi, 0.035\pi)$. This allows the coupling between the conduction band minimum and valence band maxima. In the Hatree-Fock decomposition, a finite-momentum excitonic order is found as

$$\Delta_{exc2} = -\frac{1}{N}\sum_{\boldsymbol{k}'} W(\boldsymbol{k}-\boldsymbol{k}' \pm \boldsymbol{Q})\,\langle c^{\dagger}_{\boldsymbol{k}'\mp\boldsymbol{Q},c}\, c_{\boldsymbol{k}',v}\rangle. \tag{15}$$

This order couples electrons and holes with a finite momentum difference $\pm\boldsymbol{Q}$. In the calculation, we split the Brillouin zone into small rectangular pieces with an area of $Q_x Q_y$ and consider eight pieces with the lowest energies. For simplicity, we assume an approximately constant interaction strength, $U_{CDW} = W(\boldsymbol{Q})$, between the conduction and valence bands, originating from the next-nearest-neighbor pieces in momentum space. We used the same values of $V = 1.16$ eV and $\alpha = 0.3$ used for the primary excitons. Note that the translational symmetry breaking ordering wavevector is not necessarily commensurate. The results are the same if more pieces of the Brillouin zone are considered, because the finite momentum excitonic order occurs only between the pieces that are closest to the band gap. In Fig. 5**i**, we calculated the finite momentum excitonic order parameter $\Delta_{\mathbf{exc2}}$ (blue dashed curve) and bulk energy gap $E_{\text{gap}}$ (red curve) as functions of $T$. A secondary excitonic order is developed, but it is significantly weaker than the primary one and exhibits a lower critical temperature ($T_{c2}$~25 K). It is also important to note that as the temperature decreases, the primary excitonic order develops first, opening a global bulk gap. This bulk gap would make it harder for the second finite-momentum excitonic order to form, thereby further lowering $T_{c2}$.

Note that the Coulomb interaction can, in principle, generate an infinite number of orders, each associated to a given $q$. Based on the unique band structure and experimental observations, we consider two of these orders that are the most prominent and relevant to the experiments. The first one corresponds to the $\boldsymbol{q} = \boldsymbol{k} - \boldsymbol{k}'$ component of the Coulomb interaction, resulting in a $Q = 0$ excitonic order. The second order, corresponding to $\boldsymbol{q} = \boldsymbol{k} - \boldsymbol{k}' + \boldsymbol{Q}$, results in a $Q \neq 0$ excitonic order. It is important to note that the second effect fundamentally differs from, and is not included in, the first, even though both are rooted in the same Coulomb interaction.

**Magnetic field dependence:** Here we investigate the magnetic field influence on the translational symmetry breaking ordering wavevector, based on the Zeeman coupling. As previously discussed, the wavevector can be determined by the momentum separation $\boldsymbol{Q}$ of the conduction band minimum and valence band maxima in momentum space. In the model, the magnetic field $B$ can be incorporated through a Zeeman term

$$H_{\text{Zeeman}} = h\sigma_z. \tag{16}$$

Here $h = g\mu_B B$ represents the Zeeman energy, where $g$ is the effective $g$-factor and $\mu_B$ is the Bohr magnon. As shown before, the conduction band minimum is always located at the $M$-point, and the $x$-component of $\boldsymbol{Q}$ remains relatively fixed at π, particularly in the presence of the primary excitonic order. Thus, it suffices to focus solely on the $y$-component of the translational symmetry breaking ordering wavevector. The problem can then be reformulated to investigate how $h$ influences the $y$-component of the momentum separation ($Q_y$). Its solution is illustrated by Fig. 5**k**.

The impact of magnetic fields on the band structure is complex, as it encompasses both Zeeman and orbital effects. In general, distinct orbitals generally respond differently to an applied magnetic field due to their distinct angular momenta. To simplify and illustrate the key physics, we modeled this response using an effective $g$-factor between the two orbitals, resulting in a Zeeman-like term in Eq. (16).

Considering only the role of the Zeeman effect, the gap between conduction and valence bands would be expected to exhibit two independent spin-polarized components, reducing the overall gap size as the field increases. However, additional factors such as spin-orbit coupling and the orbital coupling of the magnetic field to the electrons can change this behavior. Orbital coupling, for instance, would lead to the opposite effect of the Zeeman effect by increasing the overall gap due to the zero-point energy offset of the Landau bands or Landau levels, compared to the bare band minima/maxima[75]. A complete model explaining the observed field-induced increase in the excitonic insulator gap would have to consider the $k$-dependence of the spin textures of the gapped bands induced by spin-orbit coupling, and the role of the Zeeman effect on them. Future investigations will be required to provide a greater understanding on the field-induced increase of the excitonic insulator gap observed in $Ta_2Pd_3Te_5$.

Finally, it is important to emphasize that, from a broad perspective, the excitonic insulator is expected to endure high magnetic fields due to the absence of a Meissner effect (intrinsic to superconductors). Fenton[75] investigated the field dependence of the excitonic insulator based on orbital coupling, demonstrating that in systems with band overlap (such as $Ta_2Pd_3Te_5$), or in small bandgap semiconductors, the application of a magnetic field induces a stronger excitonic insulator than the one observed in absence of the field. Note that ref. [75] assumes isotropic bands, thus implying no anticipated shift in the wavevector. In a more complete consideration, the combination of orbital and Zeeman effects as well as Fermi surface anisotropies need to be considered. In materials featuring anisotropic bands like $Ta_2Pd_3Te_5$, this would generically also lead to a shift in the wavevector of the excitonic order under applied magnetic fields, particularly if the initial state was incommensurate. These basic trends – magnetic field enhancement of the exciton gap as well as a shift of the ordering wave vector – are fully consistent with our experimental observations.

## Methods only references

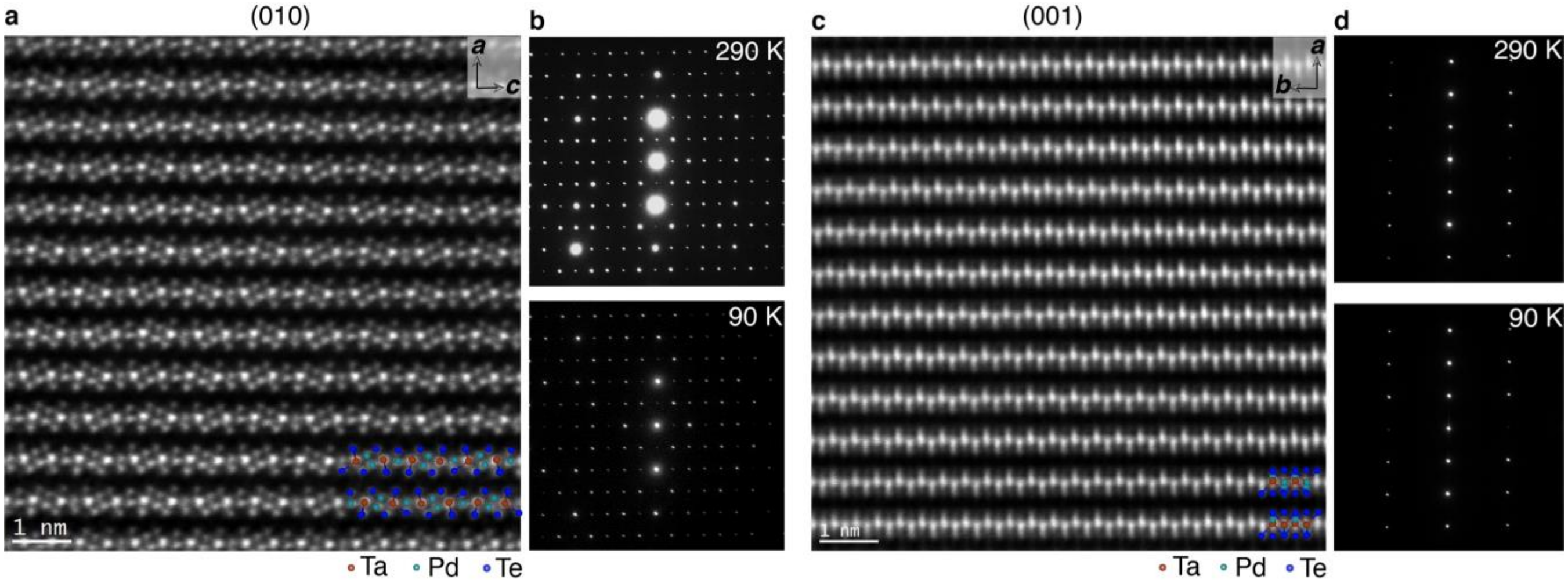

**Extended Fig. 1: Selected area electron diffraction patterns detecting no lattice instabilities near *T* = 100 K.** **a**, Atomic-resolution image of the (010) surface of a lamella captured by scanning transmission electron microscopy, demonstrating a consistent atomic arrangement when compared to pristine $Ta_2Pd_3Te_5$ (010) crystallographic structure. **b**, Selected area electron diffraction patterns obtained from the focused-ion-beam cut lamella at $T$ = 290 K (top) and 90 K (bottom). The patterns exhibit identical crystal lattice at both temperatures, detecting no structural phase change in this temperature range. **c**, Atomic-resolution scanning transmission electron microscopy image of the (001) surface, revealing a consistent atom arrangement when compared to pristine $Ta_2Pd_3Te_5$ (001). **d**, Selected area electron diffraction patterns acquired from the focused-ion-beam cut lamella at $T$ = 290 K (top) and 90 K (bottom). Akin to panel **b**, the patterns demonstrate the same crystal lattice at both temperatures. The temperature-dependent selected area electron diffraction patterns detect no structural phase transition around 100 K.

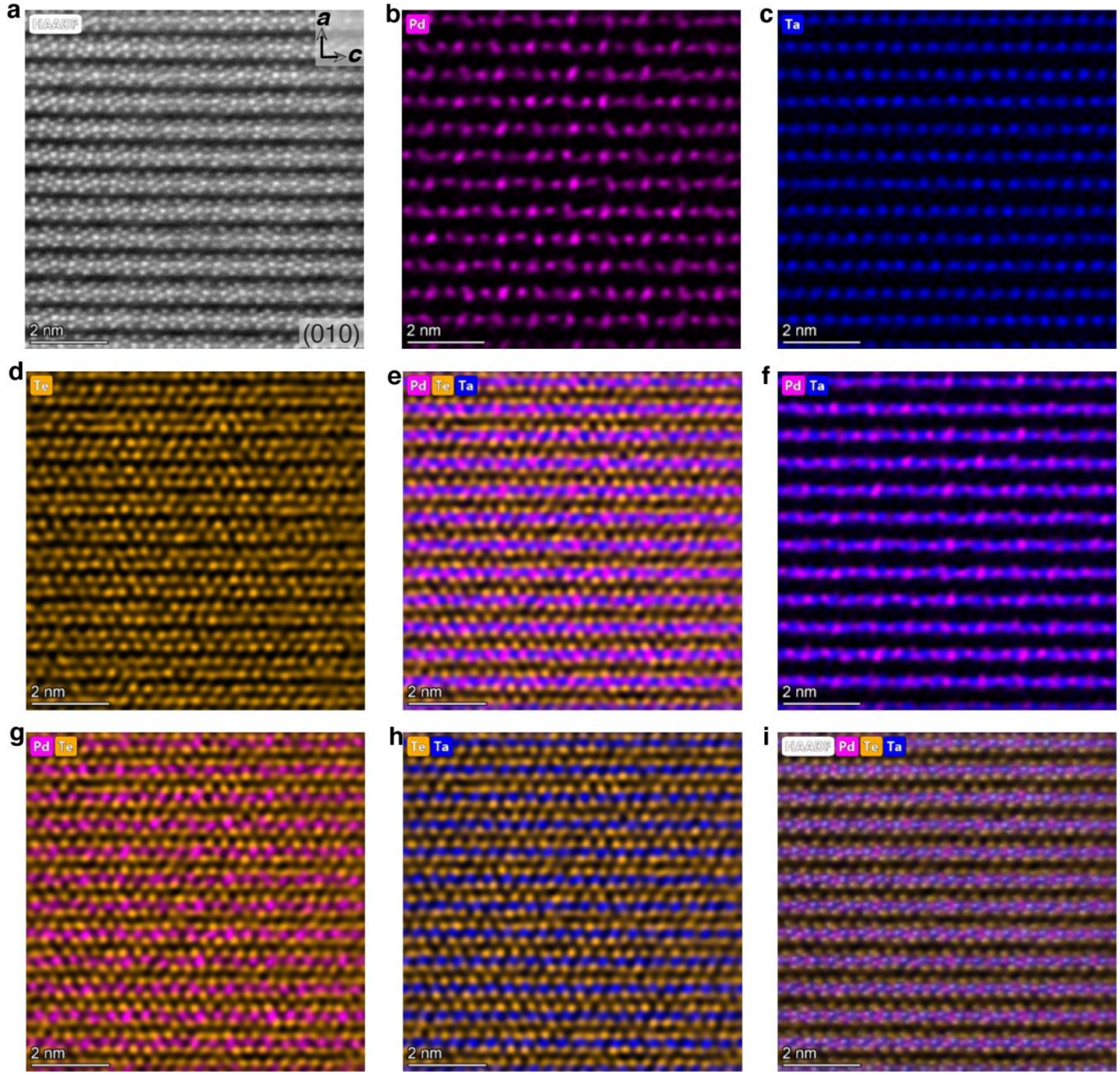

**Extended Fig. 2: Structure and phase characterization of $Ta_2Pd_3Te_5$. a,** Atomic resolution image of the side surface of $Ta_2Pd_3Te_5$ (*i.e.*, 010 plane), obtained through scanning transmission electron microscopy, displaying a consistent atomic arrangement when compared to pristine $Ta_2Pd_3Te_5$. **b-i**, Elemental mappings of the side surface utilizing an energy dispersive X-ray detector. The mappings demonstrate the unperturbed atomic layers, highlighting the structural integrity and composition of $Ta_2Pd_3Te_5$.

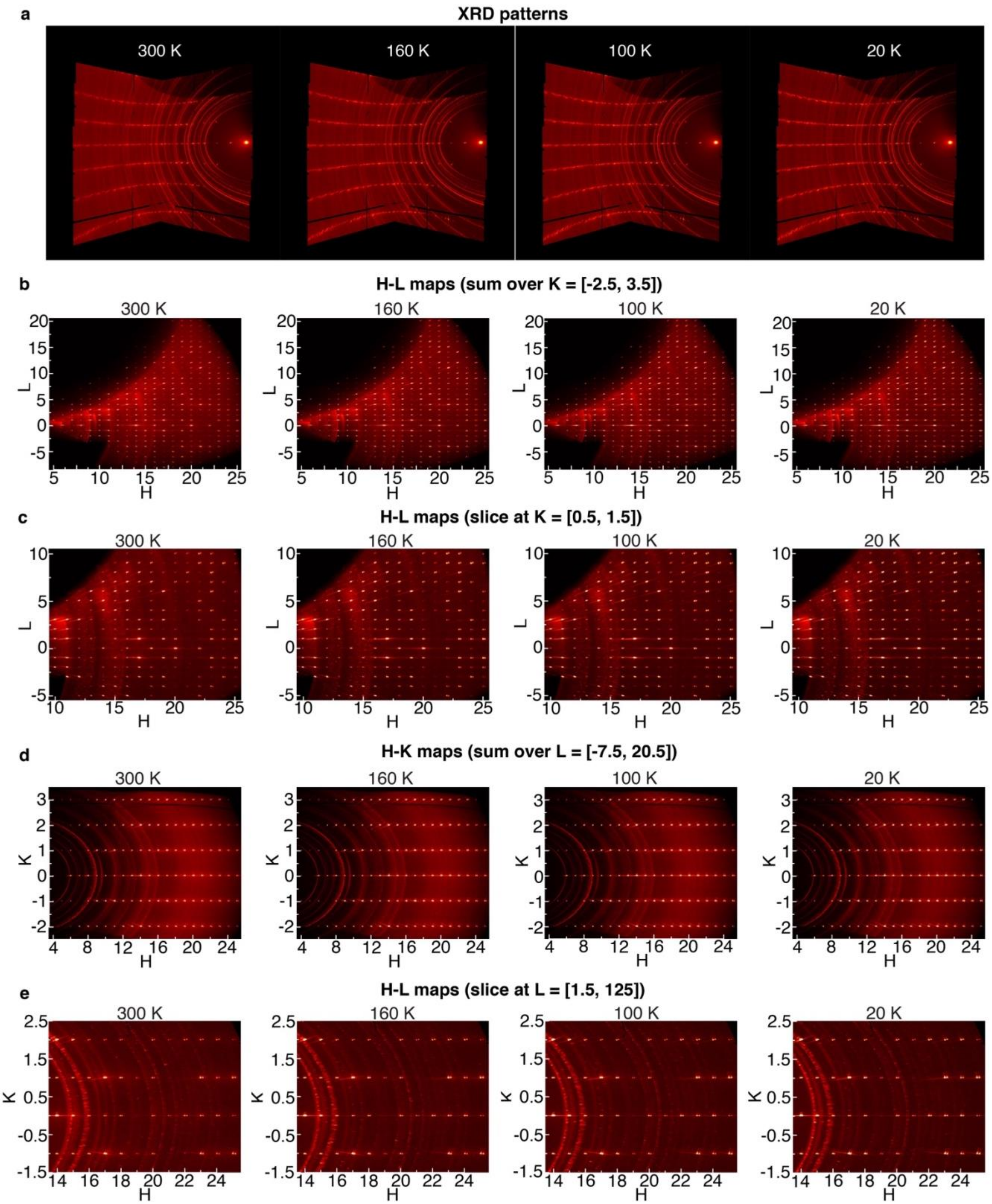

**Extended Fig. 3: X-ray diffraction measurements on $Ta_2Pd_3Te_5$ detecting no structural phase transition as a function of temperature. a,** X-ray diffraction pattern measured at the same sample position for different temperatures ranging from 300 K to 20 K. **b**, Maps along the H-L direction obtained by summing over K= [-2.5, 3.5] and measured at four temperatures. **c**, Maps along H-L obtained by taking a slice at K = [0.5,1.5] measured at four temperatures. **d**, Maps along H-K obtained by summing over L = [-7.5, 20.5] and measured at four temperatures. **e**, Maps along H-K obtained by taking a slice at L = [1.5, 2.5]. Notably, there is no emergence of any new or additional peak at low temperatures. X-ray diffraction measurements consistently exhibit a good agreement with the *Pnma* crystal structure at all the measured temperatures.

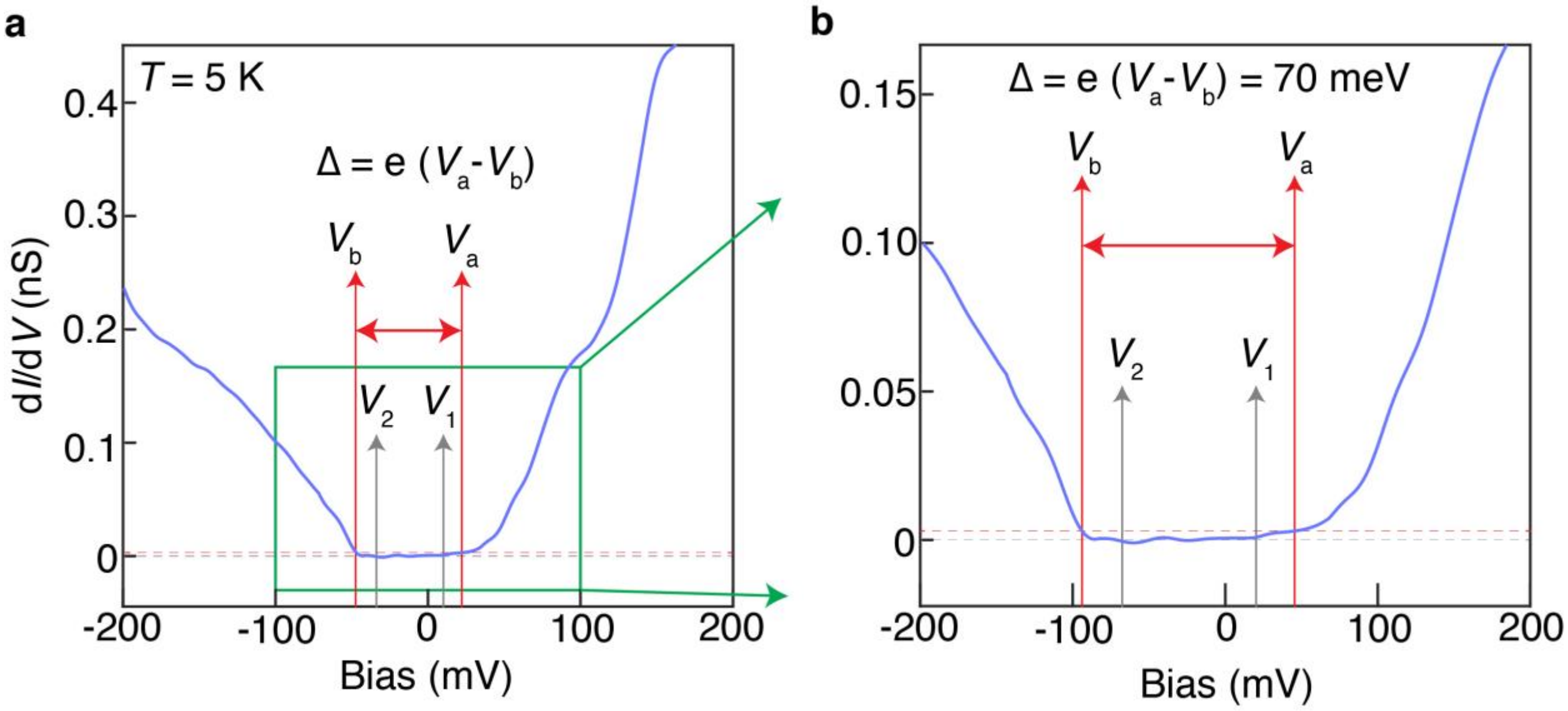

**Extended Fig. 4: Determination of the energy gap from tunneling spectroscopy. a**, Averaged d$I$/d$V$ spectrum acquired by tunneling into the *bc* plane of clean $Ta_2Pd_3Te_5$ at $T = 5$ K. The voltage interval $[V_1, V_2]$ is used to calculate the noise floor, $\sigma$. The dashed red line represents $\Gamma = 2.36\sigma$, which corresponds to the instrumental resolution of the d$I$/d$V$ signal. $V_a$ and $V_b$ are the solutions of the equation $\mathrm{d}I/\mathrm{d}V = \Gamma$. The spectroscopic energy gap is calculated as $\Delta = eV_a - eV_b$, resulting in $\Delta \simeq 70$ meV. **b**, Magnified view of panel **a**, focusing on the gapped region in the averaged d$I$/d$V$ spectrum.

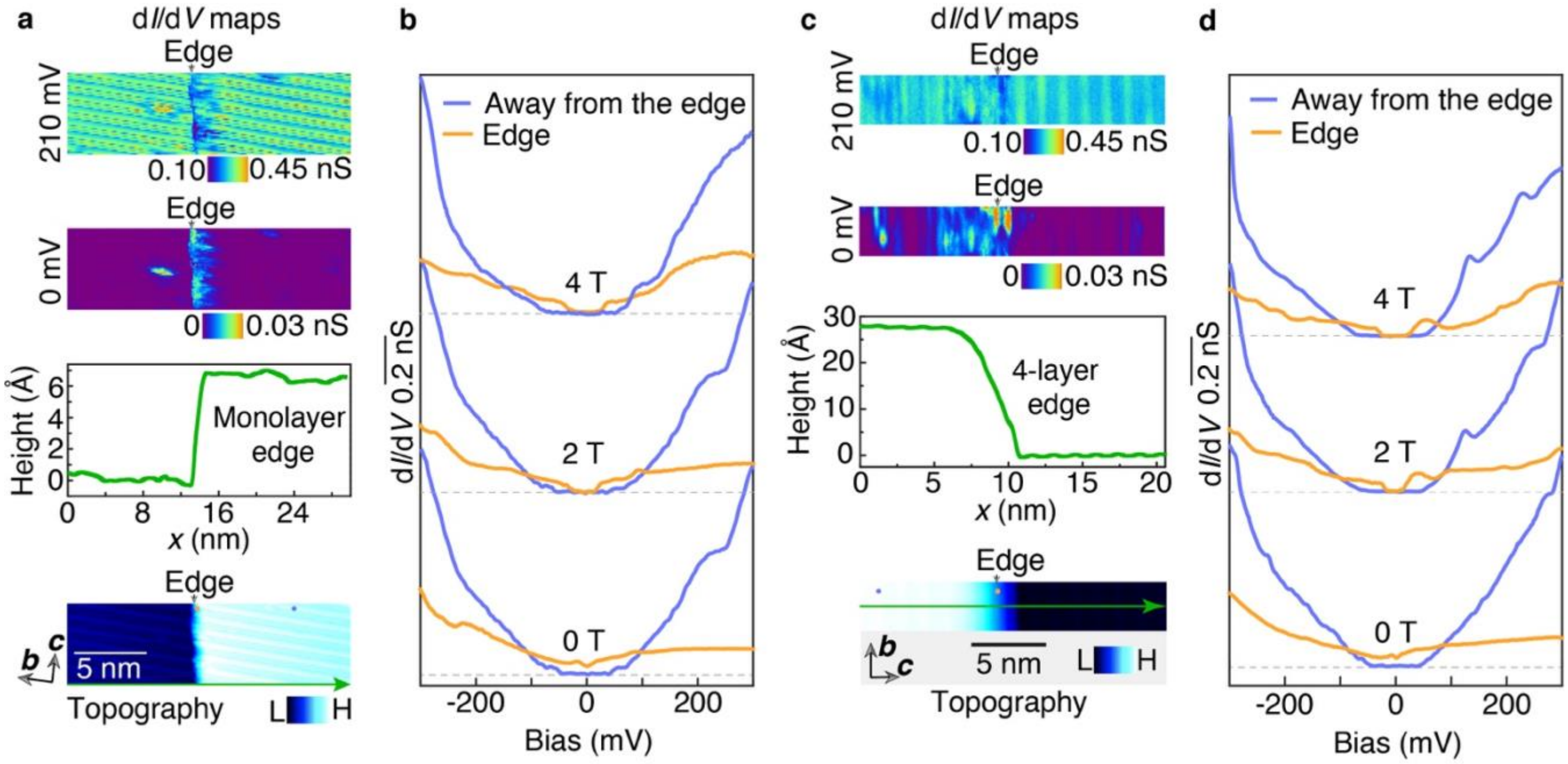

**Extended Fig. 5: Observation of time-reversal-symmetry-protected edge states for different edge configurations.** **a**, d$I$/d$V$ maps acquired at different bias voltages (corresponding topography is shown in the bottom panel) around a monolayer step edge parallel to the $c$-axis measured at $T = 5$ K. The height profile perpendicular to the $c$-axis is also displayed. The d$I$/d$V$ map obtained within the energy gap ($V = 0$ mV) reveals a pronounced edge state, whereas at $V = 210$ mV, the edge state is suppressed. **b**, Tunneling spectra acquired at locations away from the step edge and on the step edge measured under various magnetic fields. The orange and violet curves represent the differential spectra obtained at the step edge and away from it, respectively. The corresponding spatial locations where the spectra are acquired are marked with color-coded dots on the topographic image in panel **a.** Spectra at different magnetic fields were collected at the same locations and are vertically offset for clarity. Dashed horizontal lines mark the zero d$I$/d$V$ for different fields. At $B = 0$, the tunneling spectra reveal an energy gap around the Fermi energy away from the step edge, while a pronounced gapless, in-gap state is observed on the step edge. In contrast, at $B = 2$ T and 4 T, the step edge state is suppressed, and an energy gap gradually develops with increasing magnetic field. **c**, d$I$/d$V$ maps acquired at different bias voltages (corresponding topography is shown in the bottom panel) around a four-layer step edge parallel to the $b$-axis direction. The height profile perpendicular to the $b$-axis is also shown. The d$I$/d$V$ map obtained within the energy gap ($V = 0$ mV) reveals a pronounced edge state, whereas at $V = 210$ mV, the edge state is suppressed. **d**, Tunneling spectra acquired at locations away from the step edge and on the step edge measured under various magnetic fields. The orange and violet curves represent the differential spectra taken at the step edge and away from it, respectively. The corresponding spatial locations where the spectra are acquired are marked with color-coded dots on the topographic image in panel **a.** Spectra at different magnetic fields were taken at the same locations and are vertically offset for clarity. Dashed horizontal lines mark the zero d$I$/d$V$ for different fields. At $B = 0$, the tunneling spectra reveal an energy gap around the Fermi energy away from the step edge, while a pronounced gapless, in-gap state is observed on the step edge. In contrast, at $B = 2$ T and 4 T, the step edge state is suppressed, and an energy gap gradually develops with increasing magnetic field. Tunneling junction set-up: $V_{set} = 300$ mV, $I_{set} = 0.5$ nA, $V_{mod} = 2$ mV.

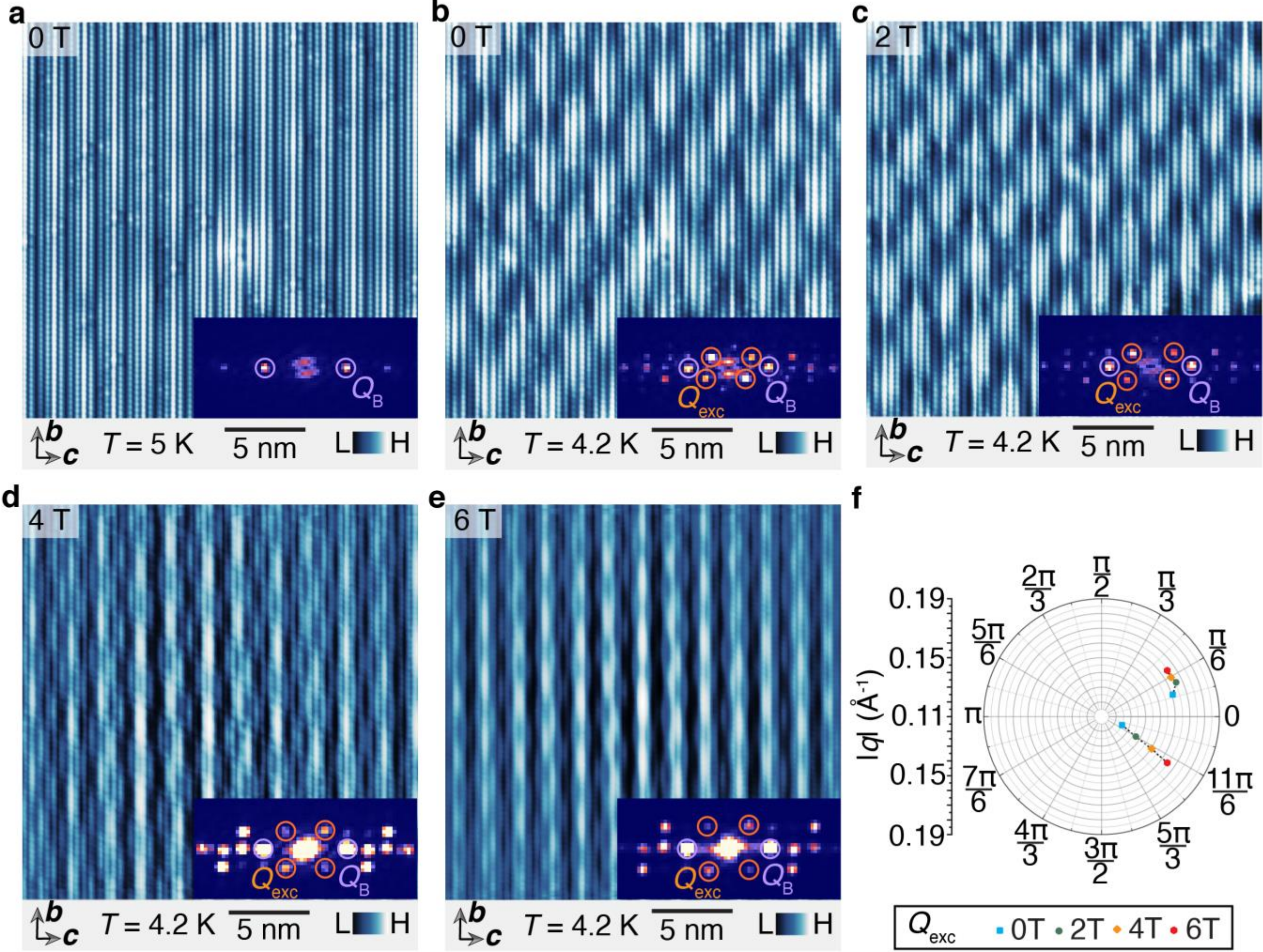

**Extended Fig. 6: Magnetic field tunability of the wavevector of the translational symmetry breaking order. a,** Atomically resolved topographic image of a clean $Ta_2Pd_3Te_5$ (100) surface acquired at $T = 5$ K. Inset shows the corresponding Fourier transform image displaying well-defined Bragg peaks (purple circles). **b**, Topographic image of the same region presented in panel **a**, acquired at $T = 4.2$ K and $B = 0$ T, revealing a pronounced translational symmetry breaking order. Inset: Fourier transform image displaying well-defined superlattice peaks (orange circles) alongside the Bragg peaks (purple circles). The extracted wavevector is $Q_{exc} = [\pm(-0.43c^* + 0.035b^*), \pm(0.57c^* + 0.035b^*)]$. **c**, **d**, **e**, Topographic images from the same location but acquired at magnetic fields of 2 T, 4 T, and 6 T, respectively, highlighting a change in the translational symmetry breaking ordering pattern upon increasing the magnetic field. The Fourier transform images shown in the inset reveal a gradually evolving $Q_{exc}$ where $Q_{exc}$ changes to $[\pm(-0.44c^* + 0.051b^*), \pm(0.56c^* + 0.051b^*)]$ at 2 T, $[\pm(-0.47c^* + 0.06b^*), \pm(0.53c^* + 0.06b^*)]$ at 4 T, and $\pm(0.5c^* + 0.07b^*)]$ at 6 T. **f**, Polar plot summarizing the magnetic field tunability of $Q_{exc}$. Starting from being incommensurate along both $b$- and $c$-axes at $B = 0$ T, $Q_{exc}$ evolves continuously and becomes commensurate along the $c$-axis at $B = 6$ T. Tunneling junction set-up: $V_{set} = 300$ mV, $I_{set} = 0.5$ nA.

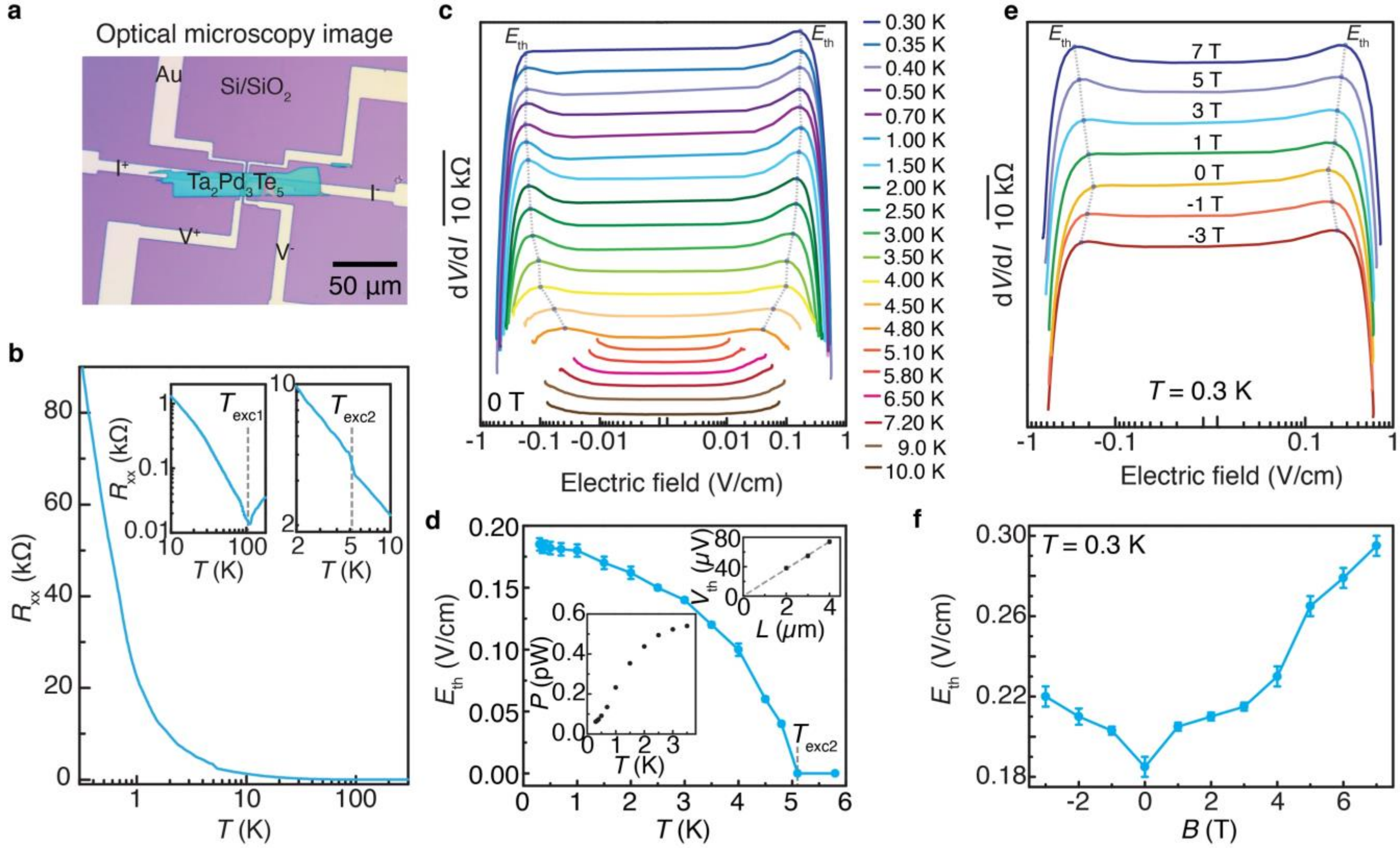

**Extended Fig. 7: Transport evidence for the translational symmetry breaking order. a**, Optical microscopy image of a four-point probe device (taken prior to its encapsulation with polymethyl methacrylate films) consisting of a mechanically exfoliated ∼100 nm thick $Ta_2Pd_3Te_5$ flake. Current and voltage leads are indicated. **b**, Four-probe resistance ($R_{xx}$) of exfoliated $Ta_2Pd_3Te_5$ as a function of the temperature, $T$. The insets display log-log plots of $R_{xx}$ ($T$) in the temperature regions near the excitonic insulator (left) and the finite momentum excitonic translational symmetry breaking (right) transitions, with corresponding transition temperatures, $T_{exc1}$ and $T_{exc2}$, marked. Below $T = 100$ K, $R_{xx}$ shows a sharp increase in resistance, indicating insulating behavior due to the formation of the excitonic insulator energy gap. Additionally, close to $T = 5$ K, $R_{xx}$ ($T$) exhibits a slight kink before continuing to increase with decreasing $T$. The onset of the translational symmetry breaking order formation is marked based on the data in panels **c** and **d**. There is no discernible hysteresis between the data corresponding to the two temperature sweep directions, *i.e.*, upon warming and cooling (see Supplementary Fig. 13). **c**, Differential resistance d$V$/d$I$ as a function of electric field at various temperatures. Traces for different temperatures are vertically offset for enhanced clarity. At low electric fields, d$V$/d$I$ remains constant for all temperatures. However, below the critical temperature $T_{exc2}$, d$V$/d$I$ sharply decreases above a threshold electric field $E_{th}$, indicating the onset of the collective phason mode or the electric field-induced sliding of the translational symmetry breaking order. ($E_{th}$ is determined as the onset of the sharp decrease in d$V$/d$I$ with respect to the electric field.) The grey dots mark the locations of $E_{th}$ for each trace, both at positive and negative electric fields. **d**, Threshold electric field $E_{th}$ as a function of temperature. The error bars represent the variation between positive and negative $E_{th}$. $T_{exc2}$ denotes the temperature at and above which $E_{th}$ becomes zero. Top inset: Threshold voltage ($V_{\mathrm{th}}$) at $T = 0.3$ K against the distance between the voltage contacts ($L$), demonstrating a linear relationship, thereby validating the relationship $E_{\mathrm{th}} = V_{\mathrm{th}}/L$ . Bottom inset: Temperature-dependent electrical power dissipation ( $P_{\mathrm{th}}$ ) at $E_{th}$, obtained from $P_{\mathrm{th}} = V_{\mathrm{th}}^2/[\mathrm{d}V/\mathrm{d}I(V_{\mathrm{th}})]$ ,

demonstrating an increase in $P_{\mathrm{th}}$ with rising temperature, indicating that the non-linear transport is not thermally driven, thus supporting the intrinsic nature of the observed properties in $Ta_2Pd_3Te_5$. **e**, d*V*/d*I* as a function of electric field at various magnetic fields. Traces for different magnetic fields are vertically offset for enhanced visibility. $E_{\mathrm{th}}$ (marked for each trace) changes with the magnetic field. **f**, Magnetic field dependence of $E_{\mathrm{th}}$. The error bars denote the variation between positive and negative $E_{\mathrm{th}}$. An external magnetic field is known to suppress the charge orders stemming from translational symmetry breaking[68,69] with the threshold electric field $E_{\mathrm{th}}$ being proportional to the amplitude of the translational symmetry breaking order parameter[70]. Therefore, the contrasting increase of $E_{\mathrm{th}}$ as a function of $B$ can be understood as resulting from the progressive, field-induced commensurability of $Q_{\mathrm{exc}}$ with the lattice constant along the *c*-axis (Fig. 4).

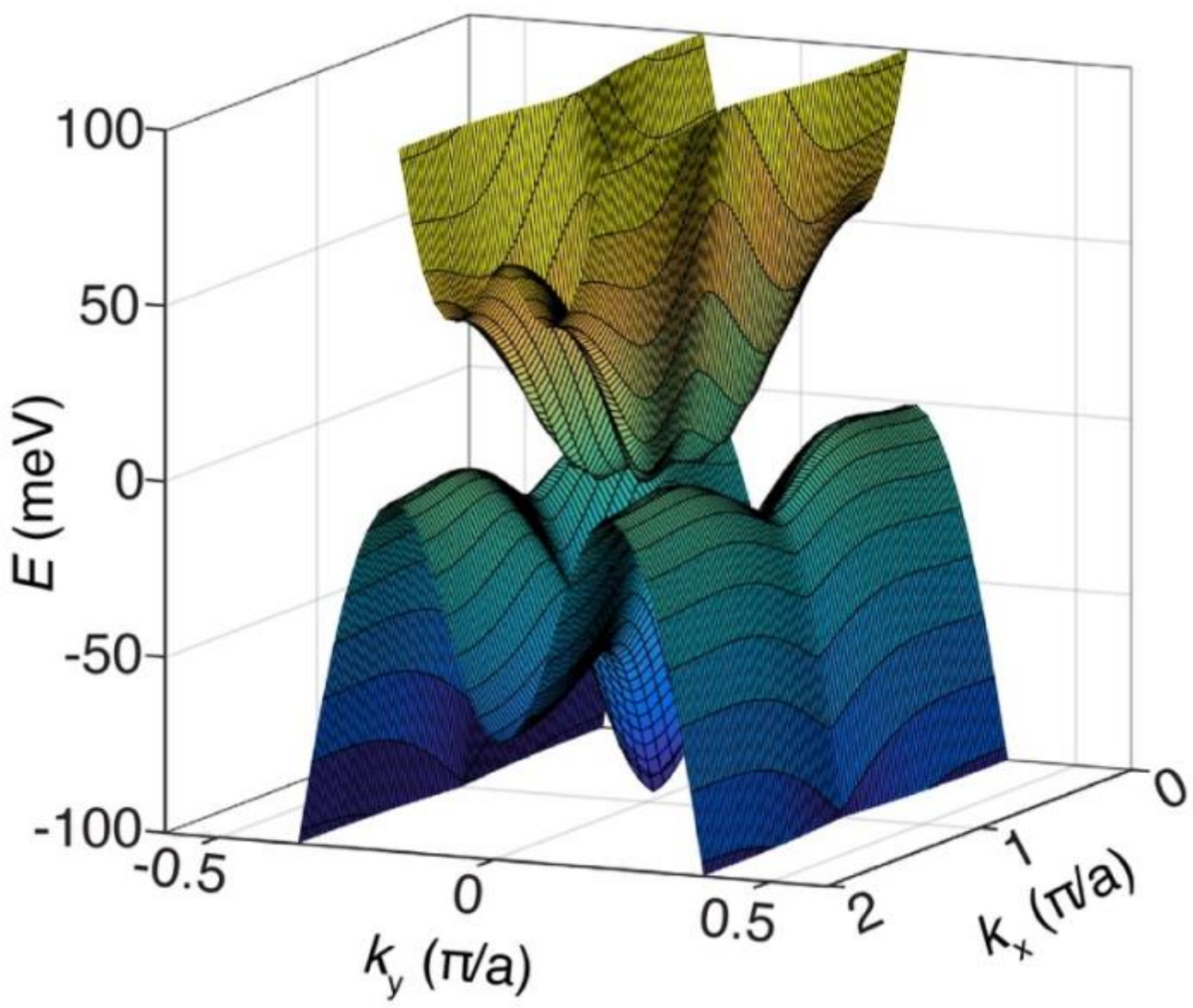

**Extended Fig. 8: Low-energy band structure of monolayer $Ta_2Pd_3Te_5$ obtained from first-principles calculations.**

**Competing interests:** The authors declare no competing interests.

**Data and materials availability:** All data needed to evaluate the conclusions in the paper are present in the paper. Additional data are available from the corresponding authors upon reasonable request.

**Acknowledgement:** Experimental and theoretical work at Princeton University was supported by the Gordon and 286 Betty Moore Foundation (GBMF4547; M.Z.H.). The material characterization is supported by the United States 287 Department of Energy (US DOE) under the Basic Energy Sciences program (grant number DOE/BES DE-FG-288 02-05ER46200). Crystal growth at Beijing Institute of Technology is supported by the National Key Research and Development Program of China (grant nos 2020YFA0308800 and 2022YFA1403400), the National Science Foundation of China (grant no 92065109), and the Beijing Natural Science Foundation (grant nos Z210006 and Z190006). L.B. is supported by DOE-BES through award DE-SC0002613. The National High Magnetic Field Laboratory (NHMFL) acknowledges support from the US-NSF Cooperative agreement Grant number DMR-DMR-2128556 and the state of Florida. We thank T. Murphy, G. Jones, L. Jiao, and R. Nowell at NHMFL for technical support. T.N. acknowledges supports from the European Union's Horizon 2020 research and innovation programme (ERC-StG-Neupert-757867-PARATOP). Z.W. thanks the Analysis and Testing Center at BIT for assistance in facility support. Y.Y.P. is grateful for financial support from the National Natural Science Foundation of China (11974029). This research used resources of the Advanced Light Source, which is a DOE Office of Science User Facility under contract no. DE-AC02-05CH11231. S.B.Z. is supported by the start-up fund at HFNL, the Innovation Program for Quantum Science and Technology (Grant nos. 2021ZD0302800), and Anhui Initiative in Quantum Information Technologies (Grant nos. AHY170000).